\documentclass{article}

\usepackage[utf8]{inputenc}
\usepackage{tismir}
\usepackage{amsmath}
\usepackage{hyperref}
\usepackage{url}
\usepackage{graphicx}
\usepackage{booktabs}
\usepackage{lipsum}

\usepackage{amssymb}%http://ctan.org/pkg/amssymb
\usepackage{ifpdf}

\makeatletter
\let\orig@endnotemark\@endnotemark
\renewcommand{\@endnotemark}{%
  \leavevmode
  \hyperlink{endnote.\theenmark}{\orig@endnotemark}%
}

\let\orig@endnotetext\@endnotetext
\renewcommand{\@endnotetext}[1]{%
  \orig@endnotetext{%
    \Hy@raisedlink{%
      \hypertarget{endnote.\theenmark}{}%
    }%
    #1%
  }%
}
\makeatother

\title{PianoCoRe: Combined and Refined Piano MIDI Dataset}
\author{%
Ilya Borovik\thanks{Skolkovo Institute of Science and Technology, Moscow, Russia}}

\date{}

\type{dataset}

\authorref{Borovik,~I.}

\journalyear{2026}
\journalvolume{9}
\journalissue{1}
\journalpages{144--163}
\doi{10.5334/tismir.333}

\authorshort{Borovik,~I.} %or, e.g., \authorshort{Author1 et al}

\begin{document}

%%%%%%%%%%%%%%%%%%%%%%%%%%%%%%%%%%%%%%%%%%%%%%%%%%%%%%%%%%%%%%%%%%%%%%%%%%%%%%%%
% Abstract
%%%%%%%%%%%%%%%%%%%%%%%%%%%%%%%%%%%%%%%%%%%%%%%%%%%%%%%%%%%%%%%%%%%%%%%%%%%%%%%%

\twocolumn[{%
\maketitleblock
\begin{abstract}
Symbolic music datasets with matched scores and performances are essential for many music information retrieval (MIR) tasks. Yet, existing resources often cover a narrow range of composers, lack performance variety, omit note-level alignments, or use inconsistent naming formats. This work presents \textbf{PianoCoRe}, a large-scale piano MIDI dataset that unifies and refines major open-source piano corpora. The dataset contains 250,046 performances of 5,625 pieces written by 483 composers, totaling 21,763 h of performed music. PianoCoRe is released in tiered subsets to support different applications: from large-scale analysis and pre-training (\textbf{PianoCoRe-C} and deduplicated \textbf{PianoCoRe-B}) to expressive performance modeling with note-level score alignment (\textbf{PianoCoRe-A/A*}). The note-aligned subset, \textbf{PianoCoRe-A}, provides the largest open-source collection of 157,207 performances aligned to 1,591 scores to date. In addition to the dataset, the contributions are: (1) a MIDI quality classifier for detecting corrupted and score-like transcriptions and (2) RAScoP, an alignment refinement pipeline that cleans temporal alignment errors and interpolates missing notes. The analysis shows that the refinement reduces temporal noise and eliminates tempo outliers. Moreover, an expressive performance rendering model trained on PianoCoRe demonstrates improved robustness to unseen pieces compared to models trained on raw or smaller datasets. PianoCoRe provides a ready-to-use foundation for the next generation of expressive piano performance research.
\end{abstract}
}]
\saythanks{}

%%%%%%%%%%%%%%%%%%%%%%%%%%%%%%%%%%%%%%%%%%%%%%%%%%%%%%%%%%%%%%%%%%%%%%%%%%%%%%%%
% Main Content Start
%%%%%%%%%%%%%%%%%%%%%%%%%%%%%%%%%%%%%%%%%%%%%%%%%%%%%%%%%%%%%%%%%%%%%%%%%%%%%%%%

\section{Introduction}
\label{sec:introduction}

Musical scores and live performances are fundamental data sources for a wide range of music information retrieval (MIR) tasks. A score provides a symbolic representation of the written composition, while a performance captures a musician’s unique interpretation through variations in timing, dynamics, and articulation. Modeling the relationship between these two domains is essential for analyzing the decisions performers make to convey musical structure and emotion to an audience. Furthermore, paired score-performance data enables computational expressive performance rendering, where trained models simulate human interpretation. For all these tasks, the scale, quality, and structure of available datasets are essential.

For piano music, numerous symbolic corpora have been developed to support computational performance analysis and modeling \citep{cancino2018computational, lerch2020analysis, emerson2025multimodal}. These resources fall into two categories. The first comprises high-fidelity recordings captured from computer-monitored acoustic pianos (e.g., Yamaha Disklavier) \citep{goebl1999vienna, hashida2018crestmusepedb, hawthorne2019maestro, foscarin2020asap, hu2023batik}. The second category relies on automatic music transcription (AMT) \citep{benetos2018amt} to generate large-scale datasets from audio recordings \citep{kong2022giantmidi, zhang2022atepp, edwards2023pijama, bradshaw2025aria, lee2025gigamidi}. While recorded datasets offer unparalleled expressive detail, they are often limited in scale and stylistic diversity. Conversely, AMT-based datasets provide diversity but often contain transcription errors and lack precise note-level alignments. Furthermore, incompatible naming schemes and metadata standards make it difficult to combine datasets without risking information leakage. Together, these challenges highlight a critical gap: a lack of a unified resource that combines the scale of transcribed data with the precision of recorded performances, all aligned to scores.

This gap is addressed by \textbf{PianoCoRe}\endnote{\url{https://github.com/ilya16/PianoCoRe}}, a comprehensive dataset that combines and refines the largest open-source piano corpora of scores and performances. PianoCoRe contains 21,763 h of piano music across 250,046 performances of 5,625 pieces by 483 composers, with scores available for 75.3\% of performances. To make this data usable across diverse applications, it is released in tiered subsets: 
\begin{itemize}
    \item \textbf{PianoCoRe-C:} a complete mixed-source piano performance collection;
    \item \textbf{PianoCoRe-B:} a deduplicated and quality-assessed subset for large-scale pre-training;
    \item \textbf{PianoCoRe-A:} a subset containing performances note-aligned to scores;
    \item \textbf{PianoCoRe-A*:} a high quality subset of the best-quality performances and note-level alignments.
\end{itemize}

Unlike previous efforts, PianoCoRe focuses on legal sustainability by restricting content to works in the public domain in the European Union, ensuring it remains a stable and sound resource for the academic community. To support diverse use cases, the dataset is archived on Zenodo\endnote{\url{https://doi.org/10.5281/zenodo.19186016}} and mirrored on Hugging Face\endnote{\url{https://huggingface.co/datasets/SyMuPe/PianoCoRe}}. 

By providing an annotated dataset that is larger and cleaner than previous resources, this work lays a foundation for the development of more intelligent computational piano performance models. 

The main contributions of the work are:

\begin{enumerate}
    \item The matching and combination of existing piano MIDI corpora into a single, large-scale unified collection with verified metadata (Section~\ref{sec:dataset});

    \item A deduplication and alignment-based heuristic for MIDI quality labeling and a trained classifier for filtering corrupted and inexpressive transcriptions, enabling the creation of the curated PianoCoRe-B dataset (Section~\ref{sec:data-quality});

    \item RAScoP (Refined Alignment for Scores and Performances), a note alignment refinement pipeline that cleans timing outliers, interpolates missing notes, and synchronizes performances with scores. It has been used to produce the note-aligned PianoCoRe-A/A* subset (Section~\ref{sec:refined-alignment});

    \item An application of PianoCoRe to the task of performance rendering and a discussion of the benefits of the combined dataset for training compared to individual source datasets (Section~\ref{sec:music-performance}).
\end{enumerate}

The rest of this work is structured as follows: Section~\ref{sec:related-work} reviews relevant piano datasets. Section~\ref{sec:dataset} details the curation process for PianoCoRe. Section~\ref{sec:data-quality} introduces the MIDI quality classifier and deduplicated subset. Section~\ref{sec:refined-alignment} presents RAScoP and the note-aligned subsets. Section~\ref{sec:music-performance} evaluates PianoCoRe on expressive performance rendering. Finally, Sections~\ref{sec:limitations} and \ref{sec:conclusion} discuss limitations and conclude the work.

\section{Related Work}
\label{sec:related-work}

This section provides an overview of the most prominent piano score and performance datasets, categorized by primary data source and intended application. Table~\ref{tab:dataset-comparison} provides a summary of the datasets relevant to PianoCoRe and statistics of PianoCoRe itself.

\begin{table*}[htb]
    \centering
    \resizebox{\textwidth}{!}{%
    \begin{tabular}{l cccc cccc}
        \toprule
        \textbf{Dataset} & \textbf{Composers} & \textbf{Pieces} & \textbf{Performances} & \textbf{Hours} & \textbf{Sources} & \textbf{Scores} & \textbf{Alignments} & \textbf{Metadata}\\
        \midrule
        MAESTRO & 43 & - & 1,276 & 199 & R & no & no & P \\
        (n)ASAP & 16 & 222 & 1,067 & 92 & R & 100\% & beat/note & P \\
        \midrule
        GiantMIDI & 2,786 & 10,855 & 10,855 & 1,237 & T & no & no & S \\
        ATEPP & 25 & 1,596 & 11,742 & 1,009 & T & 43.6\% & no & P, Q\smash{$^\dagger$} \\
        Aria-MIDI & 19,021\smash{$^\ddagger$} & - & 1,186,253 & 100,629 & T & no & no & S, P\smash{$^\dagger$} \\
        PERiScoPe & 82 & 2,738 & 46,473 & 3,784 & R, T & 81.9\% & note & P\smash{$^\dagger$} \\
        \midrule
        \textbf{PianoCoRe-C} & \textbf{483} & \textbf{5,625} & \textbf{250,046} & \textbf{21,763} & \textbf{R, T} & \textbf{75.3\%} & \textbf{no} & \textbf{P\smash{$^\dagger$}} \\
        \textbf{PianoCoRe-B} & \textbf{478} & \textbf{5,591} & \textbf{214,092} & \textbf{18,757} & \textbf{R, T} & \textbf{75.0\%} & \textbf{no} & \textbf{P\smash{$^\dagger$}, D, Q} \\
        \textbf{PianoCoRe-A} & \textbf{151} & \textbf{1,591} & \textbf{157,207} & \textbf{12,509} & \textbf{R, T} & \textbf{100\%} & \textbf{note} & \textbf{P\smash{$^\dagger$}, D, Q} \\
        \textbf{PianoCoRe-A*} & \textbf{137} & \textbf{1,517} & \textbf{130,275} & \textbf{10,330} & \textbf{R, T-HQ} & \textbf{100\%} & \textbf{note} & \textbf{P\smash{$^\dagger$}, D, Q} \\
        \bottomrule
    \end{tabular}
    }
    \caption{
        Comparison of major symbolic piano performance datasets and \textbf{PianoCoRe} dataset with its tiers.
        \textbf{Sources:} R -- recorded (Disklavier/Hardware), T -- transcribed (Audio-to-MIDI), T-HQ -- transcribed labeled as high quality.
        \textbf{Metadata:} P -- performer, S -- piano solo probability, D -- deduplication flag, Q -- quality label.
        \smash{$^\dagger$}Annotations are not available for all performances.
        \smash{$^\ddagger$}Number of unique composer names computed from raw metadata, not manually verified.
    }
   \label{tab:dataset-comparison}
\end{table*}

\subsection{Recorded MIDI Performance Datasets}
\label{subsec:recorded-datasets}

One category of datasets consists of MIDI files captured directly from human performances on computer-monitored pianos (e.g., Yamaha Disklavier). These performances offer the highest fidelity of expressive detail at the symbolic level.

The \textbf{MAESTRO} dataset \citep{hawthorne2019maestro} is the most influential in this category, with over 200 h of virtuosic performances from the International Piano-e-Competition. The high-quality, time-aligned audio-MIDI pairs have made it the standard for transcription benchmarks. However, its size and diversity are modest by modern deep learning standards. 

The \textbf{ASAP} dataset \citep{foscarin2020asap} extends MAESTRO by adding musical scores and beat annotations. The dataset contains nearly 92 h of 1,067 performances from MAESTRO aligned at the beat level to 222 unique scores. Its extension, \textbf{(n)ASAP} \citep{peter2023nasap}, adds note-level alignments, making it the largest open-source recorded MIDI dataset with score-to-performance note alignments.

Several smaller curated datasets offer exceptional detail for specialized analysis tasks.
The \textbf{Batik-plays-Mozart} corpus \citep{hu2023batik} provides note-for-note alignments between professional MIDI performances of Mozart sonatas and expert-annotated scores.
\textbf{Vienna 4x22 Piano Corpus} \citep{goebl1999vienna} captures four classical music excerpts performed by 22 pianists.
\textbf{SMD} \citep{muller2011smd} provides perfectly synchronized audio and MIDI for 50 performances of 50 pieces by 11 composers.
\textbf{MazurkaBL} \citep{kosta2018mazurkabl} provides score-aligned beats, loudness, and expressive markings for 2,000 recordings of Chopin’s mazurkas.
\textbf{CrestMusePEDB} \citep{hashida2018crestmusepedb} contains 411 note-aligned performances of 35 classical pieces by 12 pianists.
While invaluable for detailed study, these datasets’ narrow scope limits their utility for training general-purpose performance models.

\subsection{Large-Scale Transcribed MIDI Datasets}
\label{subsec:transcribed-datasets}

To avoid the time-consuming process of collecting MIDI data recorded on sensor‑equipped pianos, researchers use AMT \citep{benetos2018amt} to generate large datasets from publicly available audio.

\textbf{GiantMIDI-Piano} \citep{kong2022giantmidi} was an early large-scale piano transcription effort \citep{kong2021high}, providing 1,237 h of classical piano MIDI across 10,855 pieces. The audio was sourced from performances of IMSLP repertoire downloaded from YouTube, covering compositions from a wide range of musical periods. However, GiantMIDI-Piano does not provide any musical scores, and the metadata contains duplicates and inconsistencies (see Section~\ref{subsubsec:sources-giantmidi}).

The \textbf{ATEPP} dataset \citep{zhang2022atepp} captures 11,674 performances by renowned pianists, totaling over 1,007 h of transcribed music. About half of performances have a paired score without any note-level alignment. ATEPP provides quality labels (`high quality', `low quality', `corrupted') for some of the performances. However, as analyzed in Section~\ref{subsec:data-quality}, there are unlabeled corrupted transcriptions.

\textbf{Aria-MIDI} \citep{bradshaw2025aria} greatly expands the data scale dimension, offering over 100,629 h of transcribed piano music. Data was crawled, classified as piano solo, and annotated using a large language model-guided pipeline. The size of Aria-MIDI makes it valuable for self-supervised learning. However, the dataset lacks symbolic scores and complete annotations of musical pieces.

Other notable efforts include the \textbf{SUPRA} dataset \citep{shi2019supra}, which digitized an archive of 52 h of 478 piano roll performances. In the piano jazz domain, the \textbf{PiJAMA} dataset \citep{edwards2023pijama} provides 223 h of high-quality transcriptions of 2,777 performances by 120 pianists. 

\subsection{Mixed-Source Piano Datasets}
\label{subsec:mixed-datasets}

Although the above datasets are valuable, they exist in isolation, each with different structures and metadata conventions. Mixing them directly for piano performance modeling introduces the risk of information leakage between the training and test splits.

\textbf{GigaMIDI} \citep{lee2025gigamidi} contains over 1.4 million MIDI files from diverse single- and multi-instrument sources, including ASAP, ATEPP, GiantMIDI-Piano, Vienna 4×22, SMD, and Batik-plays-Mozart. A valuable contribution is the set of heuristics for categorizing inexpressive MIDI data. However, unnormalized piece titles in GigaMIDI complicate piece-based grouping and comparison of the data.

The \textbf{PERiScoPe} dataset \citep{borovik2025symupe} represents an effort to bridge the gap between recorded and transcription-based MIDI datasets. It contains over 35,000 note-aligned score-performance pairs, matching and combining (n)ASAP and ATEPP with 2,158 h of web-collected audio transcribed to MIDI.

The described single-source and multi-source datasets face several limitations that \textbf{PianoCoRe} aims to resolve. First, collections often lack a standardized, easy-to-navigate directory structure and verified metadata, making them difficult to combine and extend. Second, datasets may pose legal risks due to the inclusion of modern, copyrighted works. Finally, MIDI transcriptions may be duplicated, corrupted, or transcribe musical score audios that provide no information for performance analysis and modeling.

\section{PianoCoRe Dataset}
\label{sec:dataset}

This section details the construction of \textbf{PianoCoRe}. It presents a methodology for processing musical scores; matching works across diverse datasets; preprocessing the source files to resolve inconsistencies; and integrating them into a unified, navigable collection. The final dataset is presented at the end of the section.

\subsection{Notation and Definitions}
\label{subsec:definitions}

The core entities and relations used throughout the manuscript and in the data collection and processing pipelines are as follows:
\begin{itemize}
    \item \textbf{Note, $n$:} a MIDI note described by its pitch $p$, onset $o$, duration $d$, and velocity $v$: $n = (p, o, d, v)$. Notes are indexed $i \in \{1, \dots, N\}$ after sorting MIDI by onset, pitch, and duration;
    
    \item \textbf{Musical score, $y$:} a sequence of $N_s$ score MIDI notes $(y_1, \dots, y_{N_s})$;
    
    \item \textbf{Performance, $x$:} a sequence of $N_p$ performance MIDI notes $(x_1, \dots, x_{N_p})$;
    
    \item \textbf{Alignment, $A$:} a sequence of score and performance notes pairs $\{(y_i, x_j): y_i \in y \cup \{\varnothing\}, x_j \in x \cup \{\varnothing\}\}$, where $a_{ij} = (y_i, \varnothing)$ indicates a missing performed note and $a_{ij} = (\varnothing, x_i)$ -- an inserted performance note. The number of matched notes (pairs with $y_i \neq \varnothing \land x_i \neq \varnothing$) is denoted as $N_m$.
\end{itemize}

The following four ratios are used to evaluate the relationship between a score and a performance:
\begin{itemize}
    \item \textbf{Note Ratio, $R_{n}$:} a ratio of the number of notes between performance and score sequences:
    \begin{equation}
        R_n = \frac{N_p}{N_s}
    \end{equation}
    Given the same musical content, note ratio identifies structural discrepancies, such as omitted repeats ($R_n \ll 1$) or transcription noise ($R_n \gg 1$);

    \item \textbf{Alignment Recall, $R_a$:} a proportion of score notes matched to the performance:
    \begin{equation}
        R_a = \frac{N_m}{N_s} \le 1
    \end{equation}
    Recall represents the ``completeness'' of the performance relative to the score;

    \item \textbf{Alignment Precision, $P_a$:} a proportion of performed notes matched to the score: 
    \begin{equation}
        P_a = \frac{N_m}{N_p} \le 1
    \end{equation}
    High precision indicates a clean performance with few noisy notes or insertions;

    \item \textbf{Adjusted Alignment Ratio, $R'_a$:} a relaxed quality metric that takes the highest of Recall (when $N_p \ge N_s$) and Precision ($N_p < N_s$):
    \begin{equation}
        R'_a = \frac{N_m}{\min(N_s, N_p)} = \max(P_a, R_a) \le 1
    \end{equation}
    It ensures that a performance is not penalized for missing notes (e.g., skipped repeats) as long as the played notes match the score, and is not penalized for extra notes (e.g., transcription noise) as long as all score notes are present.
\end{itemize}

Furthermore, the two common types of symbolic errors handled during preprocessing are:
\begin{itemize}
    \item \textbf{Duplicate Notes:} two or more notes having the exact same pitch, onset time, and duration;
    
    \item \textbf{Overlapping Notes:} a condition where a note $n_i$ of pitch $p$ starts while a previous note $n_{i-1}$ of the same pitch is still active ($o_{i} < o_{i-1} + d_{i-1}$).
\end{itemize}

\subsection{Data Matching Methodology}
\label{subsec:data-matching}

The essential part of a score and performance music dataset is the correct matching of scores and performances. One approach is to use composition entity resolution \citep{kong2022giantmidi, zhang2022atepp} that compares the titles and available metadata for score and performance files. However, the music content may not reflect the title if the file is mislabeled or has a unique naming format. 

MIDI-to-MIDI matching is used to combine datasets. This allows one to directly compare notes in musical scores and performances. It also enables one to match performances to musical scores that are only available in MIDI format and have no MusicXML \citep{good2001musicxml} counterpart. Finally, it allows one to match performances to other performances to obtain more labeled data when no scores are available.

\subsubsection{Score Processing}
\label{subsubsec:score-processing}

Before matching, the MusicXML files were converted to MIDI format using the \texttt{partitura} library \citep{cancino2022partitura} with the following refinements:
\begin{itemize}
    \item \textbf{Dynamics and Tempo:} the \texttt{<sound>} tags and \texttt{dynamics} attributes for notes are processed to embed performance direction markings for dynamics and tempo directly into the note velocities and tempo changes of the score MIDI file.

    \item \textbf{Ornaments:} trills and mordents are unrolled based on the invisible notes available in MusicXML (\texttt{<cue/>} or \texttt{print-object=``no''}). The base visible ornament note is removed to avoid overlapping note events.

    \item \textbf{Grace Notes:} acciaccatura and appoggiatura notes are expanded based on the definitions. Acciaccatura notes appear as a sequence of 32nd notes before the beat. Appoggiatura notes steal the duration of the main note.
    
    \item \textbf{Repeats:} for scores with repeats, two versions are created: a \textit{maximal} version with all repeats unfolded and a \textit{minimal} version with each repeat played only once (suffix \texttt{\_mini} in the file name). 
\end{itemize}

These changes ensure fair consideration of score structure and performance-specific elements in MIDI score files. To simplify the management of the created dataset, the full set of possible repeat structures in the scores was not considered.

\subsubsection{Candidate Pair Selection}
\label{subsubsec:candidate-pairs}

To avoid a brute-force comparison of all files, a filtering step to identify a smaller set of candidate pairs is performed. A score is paired with a performance if they meet the following criteria:
\begin{itemize}
    \item \textbf{Composer:} the composer names, extracted from file paths or metadata tags, match;
    
    \item \textbf{Note Count:} the note ratio $R_n$ falls within a plausible range of close length: $0.75 \le R_n \le 1.33$;
    
    \item \textbf{Keywords:} if available, the catalog numbers, and key/scale information within the titles match.
\end{itemize}
This pre-filtering enables efficient application of computationally intensive, alignment-based verification.

\subsubsection{Note Alignment and Verification}
\label{subsubsec:refined-alignment}

For the final step, note-level alignments for candidate pairs were computed using the \texttt{DualDTWNoteMatcher} from Parangonar \citep{peter2023parangonar}. The underlying dynamic time warping (DTW) implementation was optimized using Numba's just-in-time (JIT) compilation \citep{lam2015numba}. The optimized version works, on average, 12 times faster, on the ASAP dataset. This optimization was essential for performing millions of pairwise alignments within a reasonable timeframe.

A candidate pair is considered a definitive match if the alignment achieves $R_a > 0.7$ (more than 70\% of score notes matched to the performance). This threshold was chosen empirically to ensure a global overlap between the sequences with close score and performed repeat structures. Unmatched notes may correspond to omitted repeats, transcription errors, or specific interpretations. These are still valuable for performance-only applications, including large-scale pre-training.

Performances that fail to align with the maximal unfolded score are matched to the minimal one, increasing data retention. The exact repeat structure of the performances is not detected. For trills, the number of notes may differ between performances and scores. However, unrollment of trills in the score MIDI yields a higher alignment recall than aligning multiple performed notes to a single base trill note. 

Alignments are stored in compressed \texttt{.npz} files compatible with the original MIDI files. Each file contains arrays describing the attributes of the aligned score and performance notes: indices, pitches, and onset/offset times. Insertions and deletions are represented by the sentinel value -1 for missing attributes.

\subsection{Source Performance Datasets}
\label{subsec:performance-sources}

PianoCoRe is built by refining and integrating open-source piano MIDI datasets. This section describes the steps taken to improve the quality of source datasets before combining them under a single collection.

\subsubsection{ASAP}
\label{subsubsec:sources-asap}

The (n)ASAP dataset v2.1.1 \citep{peter2023nasap}\endnote{\url{https://github.com/CPJKU/asap-dataset}} was used. The original score MIDI files, exported using MuseScore \citep{watson2018musescore}, contain data parsing issues like unrealistic time signatures (e.g., 65/4, 25/32), cut measures with anacrusis, duplicated notes, and notes with zero duration. These were corrected by re-generating score MIDI files using the pipeline from Section~\ref{subsubsec:score-processing}. The performance MIDI files were cleaned by removing duplicate notes, truncating durations of the first of the two overlapping notes (such that $o_i = o_{i-1} + \hat{d}_{i-1}$), and removing all notes shorter than 5 ms. There are 208 score and 94 performance MIDI files with zero duration notes in the original dataset.

\subsubsection{ATEPP}
\label{subsubsec:sources-atepp}

The ATEPP v1.2 dataset \citep{zhang2022atepp}\endnote{\url{https://github.com/tangjjbetsy/ATEPP}} was used. Only 5,091 of 11,674 transcribed performances are paired with scores without an alignment. ATEPP shares the scores with ASAP but not all suitable scores (e.g., the entirety of Chopin) are present in ATEPP. By matching two datasets, 39 scores from ASAP can be assigned to 827 performances in ATEPP. 

As a preprocessing step, score MIDI files were computed from MusicXML files, similar to ASAP. Also, the following metadata issues were corrected: merging duplicate movements under different names (49 movements and 265 reassigned performances), performances with a wrong piece name (24 movements and 43 performances), and performances without a score in the metadata (3 scores and 14 performances). These problems were fixed by matching and checking performances and scores of the same composer.

\subsubsection{GiantMIDI-Piano}
\label{subsubsec:sources-giantmidi}

For GiantMIDI-Piano \citep{kong2022giantmidi}, a curated subset of the original data\endnote{\url{https://github.com/bytedance/GiantMIDI-Piano}} consisting of 7,236 MIDI files was used. The analysis of the metadata showed duplicates (by YouTube ID) in the original curated data. In total, 315 MIDI transcriptions were distributed under multiple composition names. Also, manual inspection during the matching process revealed other inconsistencies. A MIDI file may represent only a specific movement of the annotated piece, or it may be a performance of a different piece mistakenly matched after a YouTube search.

Since checking and annotating all MIDI files is exhaustive, only sequences that matched with the scores and performances from other examined datasets were used. The final subset included 2,139 performance MIDI files of musical pieces by 402 composers.

\subsubsection{PERiScoPe}
\label{subsubsec:sources-periscope}

The PERiScoPe v1.0 dataset  \citep{borovik2025symupe}\endnote{\url{https://huggingface.co/datasets/SyMuPe/PERiScoPe}} was processed by excluding performances from ASAP or ATEPP. Only the remaining 34,773 performance MIDI files transcribed from audio sources using Transkun V2 \citep{yan2024transkun} were used. The dataset required no specific process except for common transcription artifacts, described below in Section~\ref{subsubsec:transcription-errors}.

\subsubsection{Aria-MIDI}
\label{subsubsec:sources-ariamidi}

From the Aria-MIDI v1 dataset \citep{bradshaw2025aria} with 1,186,253 transcribed MIDI files\endnote{\url{https://huggingface.co/datasets/loubb/aria-midi}}, 621,132 files that had a composer in the metadata were filtered and used. There are 19,021 unique composer names in the filtered subset. 

An important difference in Aria-MIDI is how sustain pedals are encoded. The transcribed files do not distinguish between pressed and sustained note durations. The durations were predicted as sustained even when the sustain pedal was predicted separately.

\subsubsection{Transcription Artifacts}
\label{subsubsec:transcription-errors}

One issue fixed for all transcribed MIDI datasets is the error with `infinite' pitches, where notes span until the end of the file. This artifact arises when open-source transcription models \citep{kong2021high, yan2024transkun} produce unmatched note-on and note-off events due to offset or sustain-pedal decoding errors. During MIDI serialization, such notes remain active till the end of the sequence. An algorithm to identify and correct note durations was developed to repair performances in the source datasets: ATEPP (30 MIDI files), GiantMIDI (9), PERiScoPe (92), and Aria-MIDI (5,501).

\ifpdf
\begin{figure*}[ht!]
    \centering
    \includegraphics[width=1.\textwidth,alt={Flowchart of the three-stage data matching and annotation pipeline}]{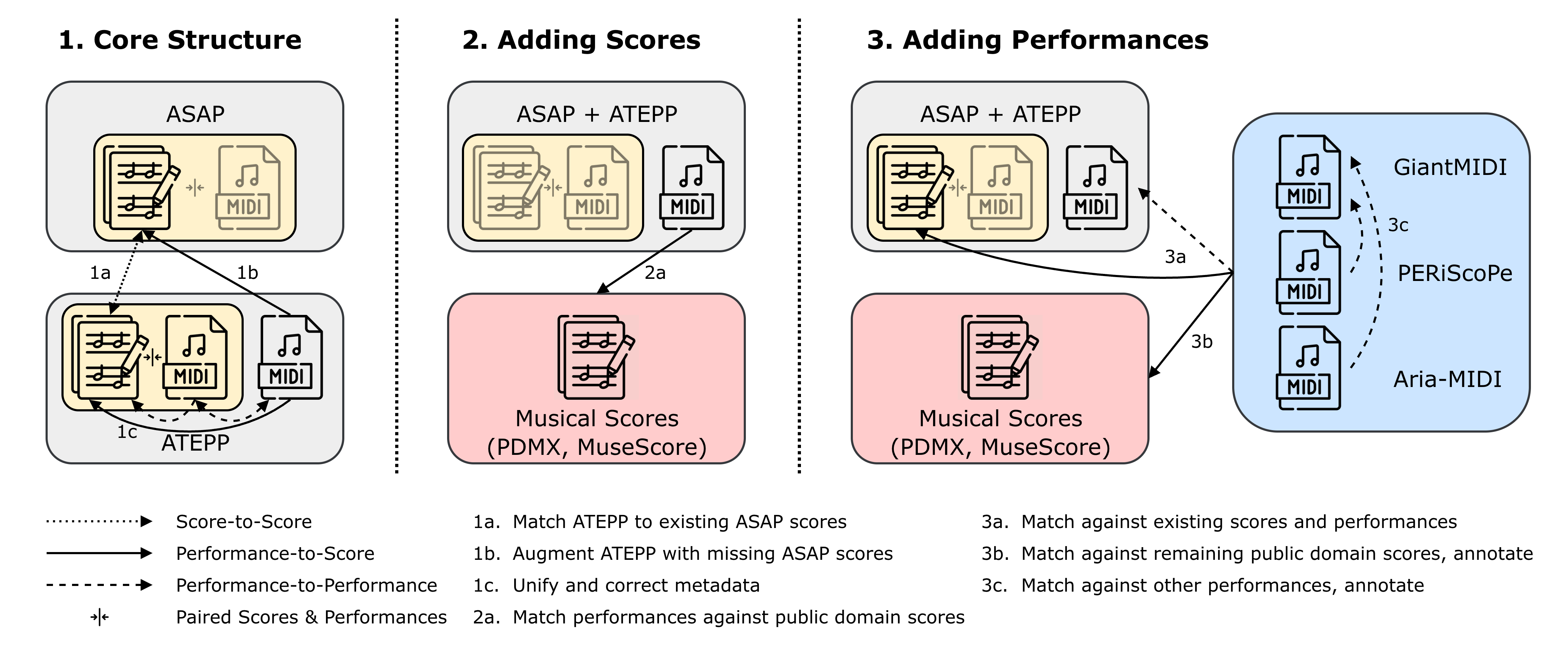}
    \caption{
        The three-stage data matching and annotation pipeline used to create PianoCoRe dataset. 
    }
    \label{fig:data-matching}
\end{figure*}
\else\fi

\subsection{Musical Score Data Sources}
\label{subsec:score-sources}

To maximize the number of aligned performances, the score library was expanded beyond ASAP and ATEPP and included public domain MusicXML scores from the PDMX dataset \citep{long2025pdmx}, originally sourced from MuseScore\endnote{\url{https://musescore.com/sheetmusic}}. In addition, the sequenced MIDI scores from KunstderFuge\endnote{\url{https://kunstderfuge.com}} and ClassicalMIDI\endnote{\url{https://www.classicalmidi.co.uk}} websites were used solely for enriching the representation of annotated performed compositions in PianoCoRe. The copyrighted scores are not redistributed in the final dataset. Since KunstderFuge provides live performance and orchestral MIDI files, inexpressive solo piano sequences were filtered out using a Note Onset Median Metric Level (NOMML) heuristic from GigaMIDI \citep{lee2025gigamidi}. Finally, during the iterative data matching process, 421 public domain scores from MuseScore were manually sourced for the most frequently performed compositions that lacked a score.

\subsection{Data Combination Process}
\label{subsec:data-combination}

The \textbf{PianoCoRe} dataset was assembled using a semi-automated, iterative process designed to merge multiple sources into a single, structurally unified collection. This process relies on the data matching and note alignment (Section~\ref{subsec:data-matching}), supplemented by manual curation and labeling to resolve ambiguities. 

The main strategy was to establish a unified data organization and gradually integrate scores and performances from source datasets. The combination process unfolds in three stages, illustrated in Figure~\ref{fig:data-matching}:

\begin{enumerate}
    \item \textbf{Core Structure:} The process began with the merging of two foundational datasets: ASAP and ATEPP. Performances and scores from ASAP were matched and reorganized into the unified ATEPP directory structure. Lastly, the 21 ASAP pieces not present in ATEPP were distributed under new directories. This created a unified base of recorded and transcribed performances with their corresponding scores.

    \item \textbf{Adding Scores:} The core dataset was then augmented by matching its performances against a large corpus of scores from PDMX, KunstderFuge (KDF), and ClassicalMIDI (CM), along with manually added MuseScore (MS) files.

    \item \textbf{Adding Performances:} The final step involved the integration of the performance datasets: GiantMIDI-Piano, PERiScoPe, and Aria-MIDI. Performances were matched against available scores based on the initial candidate pair selection. If a piece was not present in the dataset, a new directory containing the score and matched performances was added. To further increase data coverage, remaining performances were matched against those without a score from ATEPP and against each other to identify additional composition-based links.
\end{enumerate}

\ifpdf\else
\begin{figure*}[ht!]
    \centering
    \includegraphics[width=1.\textwidth,alt={Flowchart of the three-stage data matching and annotation pipeline}]{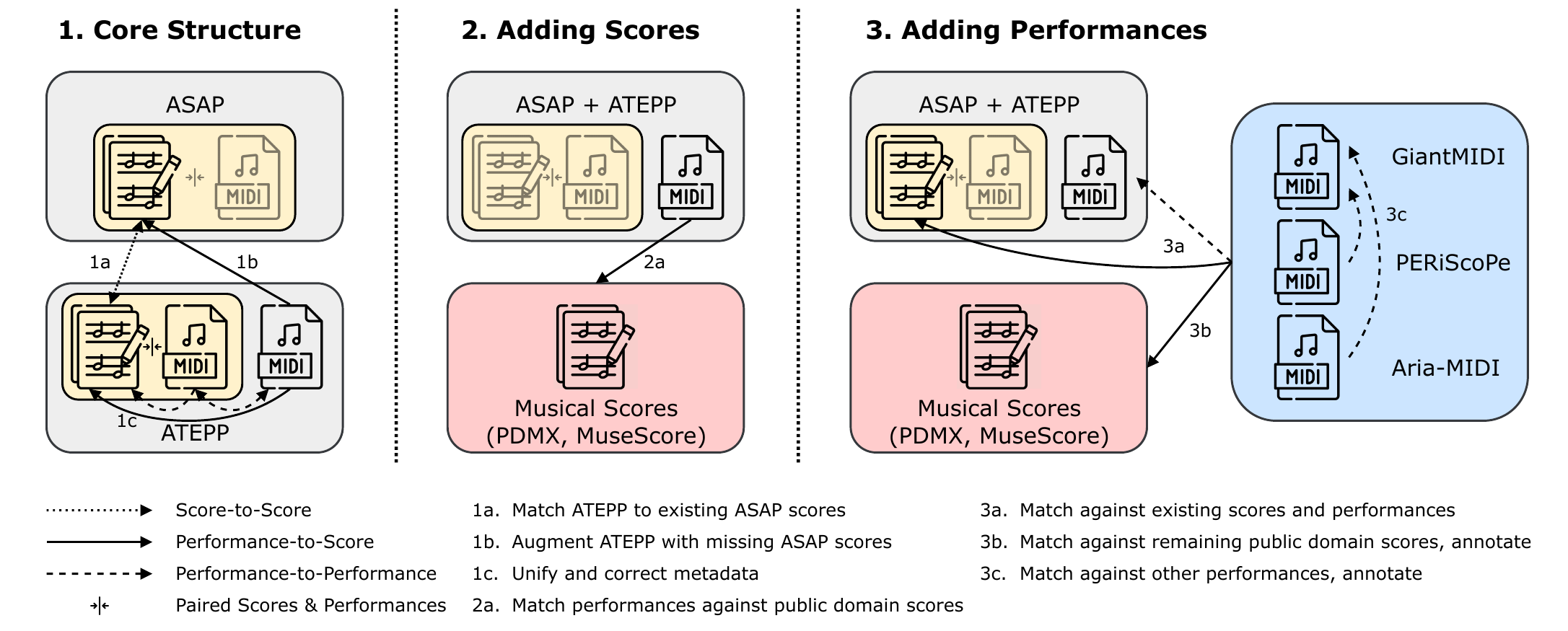}
    \caption{
        The three-stage data matching and annotation pipeline used to create PianoCoRe dataset. 
    }
    \label{fig:data-matching}
\end{figure*}
\fi

Throughout the process, automated matches were reviewed. For new pieces, composition and movement titles were manually verified and standardized using IMSLP\endnote{\url{https://imslp.org}} and web search. This step ensured consistency, corrected mislabeled files, and prevented works from being cataloged under different names. To ensure compliance with copyright standards, only works in the public domain in the European Union\endnote{\url{https://eur-lex.europa.eu/EN/legal-content/summary/copyright-and-related-rights-term-of-protection.html}} were included.

\subsection{PianoCoRe-C Dataset}
\label{subsec:pianocore-c}

\begin{figure*}[htb]
    \centering
    \includegraphics[width=1.\textwidth,alt={Coverage of musical works and performances for top-50 composers}]{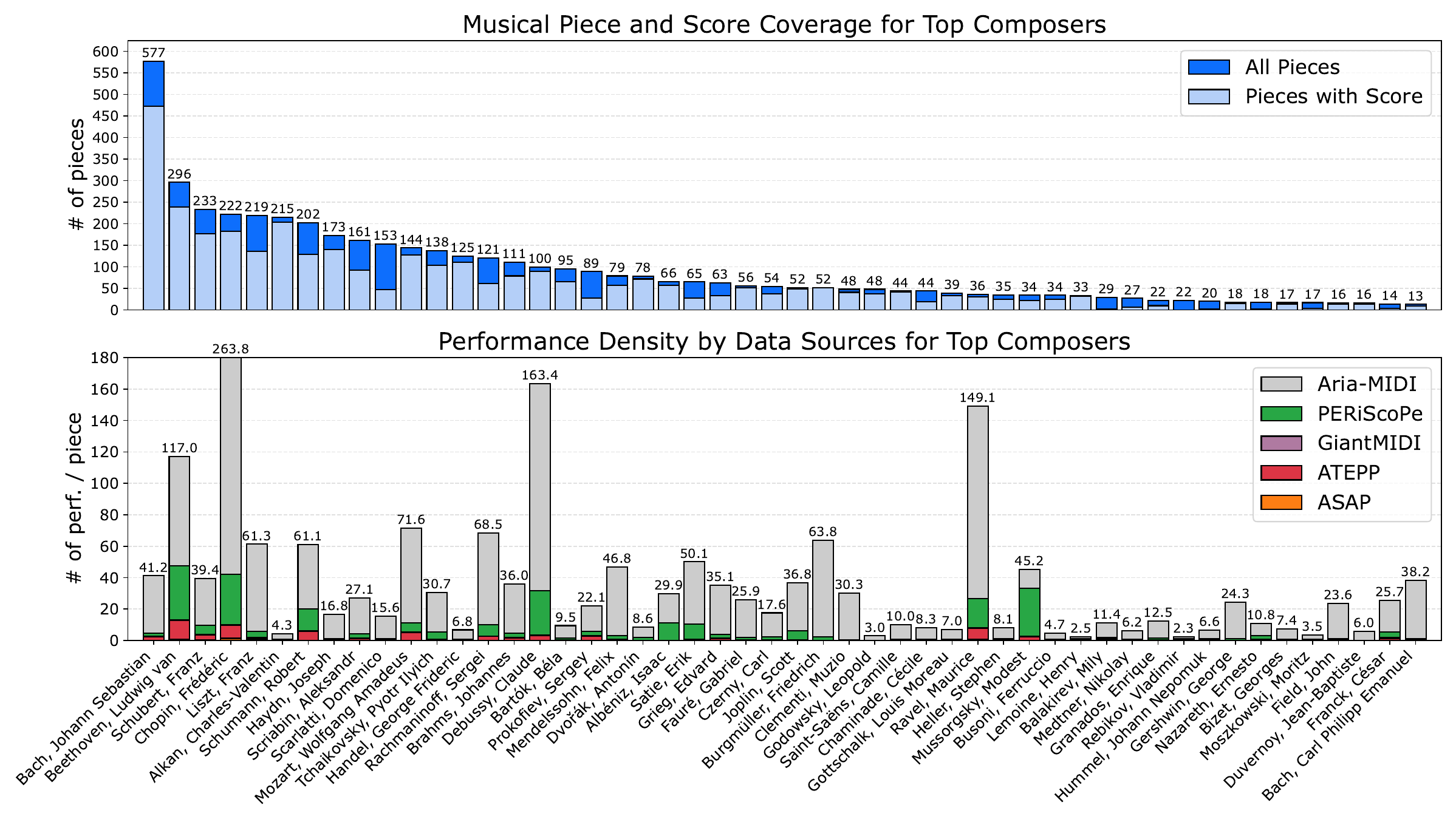}
    \caption{
        Statistical overview of the \textbf{PianoCoRe-C} dataset for the 50 most represented composers.
        \textbf{Top}: The total number of unique pieces per composer (blue) and the number of pieces with a musical score (light blue). 
        \textbf{Bottom}: The average number of performances per piece, accumulated by the MIDI source. 
    }
    \label{fig:dataset-stats-composer}
\end{figure*}

\begin{figure}[htb]
    \centering
    \includegraphics[width=1.\columnwidth,alt={Histogram showing musical pieces distributed by number of performances}]{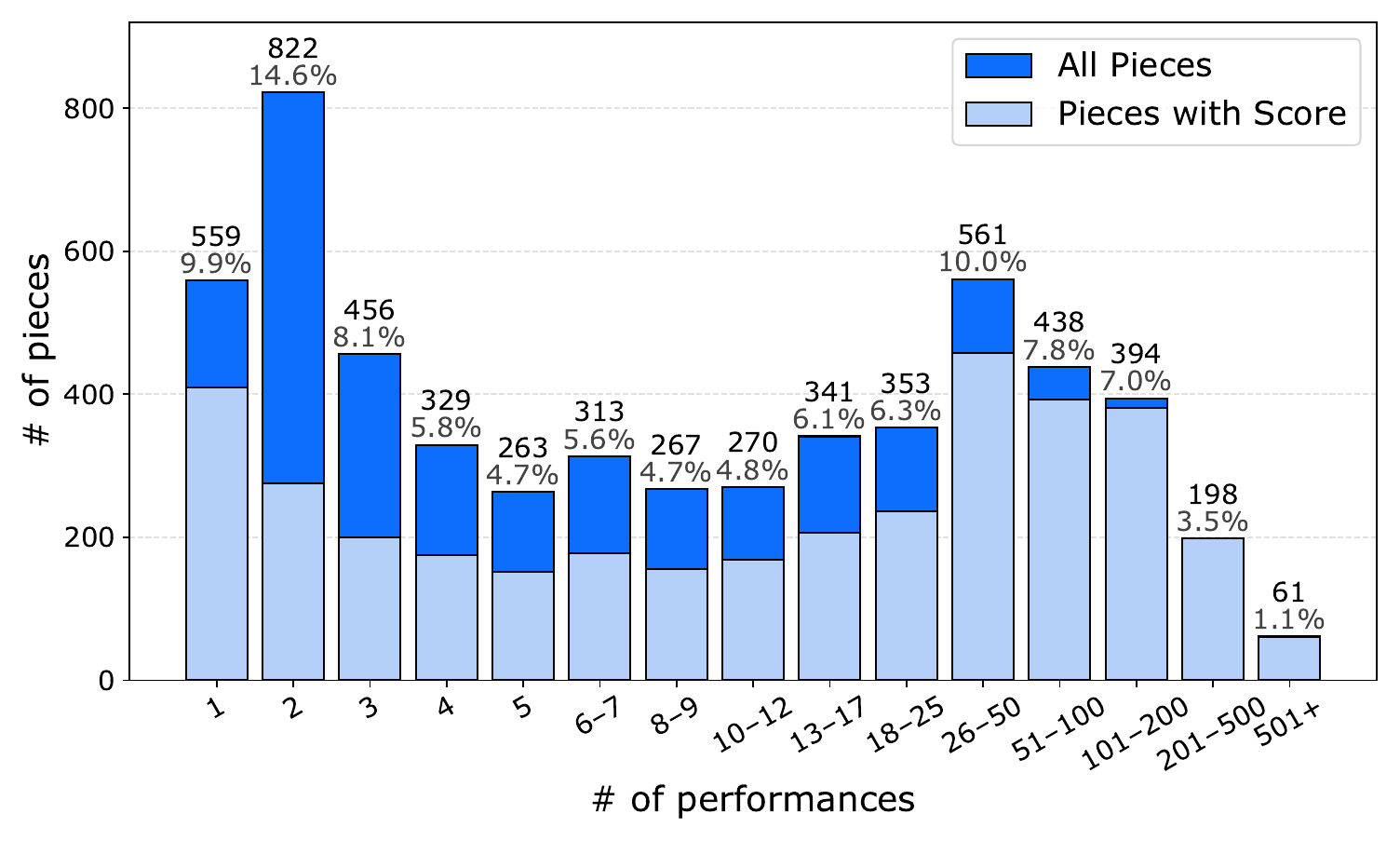}
    \caption{
        Distribution of the number of musical pieces by the number of performances in PianoCoRe-C.
    }
    \label{fig:dataset-stats-piece}
\end{figure}

The result of the data combination is \textbf{PianoCoRe-C} dataset, where `C' stands for `Core' or `Combined'. This dataset represents the most diverse collection of piece-wise annotated piano performances. It contains 250,046 performance MIDI files for piano pieces composed by 483 composers from different historical periods and styles: ranging Baroque, Classical and Romantic to Impressionist and Modern. There are 2,869 unique compositions and 5,625 unique pieces and movements. Figure~\ref{fig:dataset-stats-composer} highlights the distributions of pieces and performances per piece for popular composers. Figure~\ref{fig:dataset-stats-piece} shows the distribution of the number of musical pieces by the number of performances. The median and mean numbers of performances per piece are equal to 8 and 44, respectively. In total, 1,104 musical pieces have 50+ performances samples.

Note that \textbf{PianoCoRe-C} is not deduplicated or filtered for quality. 
This raw, comprehensive collection serves as the foundation for the refined subsets, \textbf{PianoCoRe-B} and \textbf{PianoCoRe-A}, detailed next.

\subsubsection{Content and Metadata}
\label{subsubsec:content-metadata}

All score and performance files are organized under the \texttt{composer/composition/movement/} directory hierarchy, making the dataset easy to navigate and parse. The following unified naming convention is used:
\begin{itemize}
    \item \textbf{Composer}: composer directories follow IMSLP format \texttt{[last\_name],\_[first\_name]};
    
    \item \textbf{Piece}: piece/opus numbers are represented using Arabic numbers, scales follow the format \texttt{[Note]\_[?sharp|flat]\_[major|minor]};
    
    \item \textbf{Filename}: The source of every file is preserved in the metadata and the filename, formatted as \texttt{[source]\_[original\_filename].mid}.
\end{itemize}

The dataset provides content and metadata to support various performance analysis and modeling tasks:

\begin{itemize}
    \item \textbf{Score}: MusicXML and MIDI files, source (ASAP, ATEPP, PDMX or MuseScore), note count;
    
    \item \textbf{Performance}: MIDI file, source (ASAP, ATEPP, GiantMIDI, PERiScoPe, or Aria), flag for a transcribed performance and transcription model name, performer's name (if available), duration and note count;
    
    \item \textbf{Quality Labels}: lead performance (the higher priority version of the performance for duplicates), MIDI quality class probabilities and predicted label (`score', `high quality', `low quality', or `corrupted') (Section~\ref{sec:data-quality});
    
    \item \textbf{Alignment}: if available, path to the \texttt{\_align.npz} file with raw alignment (after Parangonar), path to the \texttt{\_refined\_align.npz} file with the complete note-to-note alignment between the score and cleaned performance, and alignment recall/precision before and after alignment refinement (Section~\ref{sec:refined-alignment});
    
    \item \textbf{Refined Performance}: if alignment is available, refined MIDI file (real and synthetic notes annotated using MIDI markers) that has a complete note alignment with the score MIDI (Section~\ref{sec:refined-alignment}).
\end{itemize}

\subsubsection{Applications}
\label{subsubsec:pianocore-c-applications}

\textbf{PianoCoRe-C} includes matched score and performance MIDI files from the existing piano score and performance datasets: ASAP, ATEPP, PDMX, GiantMIDI-Piano, Aria-MIDI, and PERiScoPe. The combined dataset can be used for tasks that benefit from maximum data scale, such as self-supervised pre-training of music models, large-scale music analysis, or developing data cleaning and filtering techniques.

\section{Performance MIDI Quality Assessment}
\label{sec:data-quality}

The \textbf{PianoCoRe-C} dataset contains MIDI files of varying quality, including duplicates. This limits its application to expressive performance modeling. This section details the two-stage refinement process used to produce \textbf{PianoCoRe-B} (`B' for `Base'), a deduplicated, and quality-labeled subset of the data.

\subsection{Content-Based Performance Deduplication}
\label{subsec:deduplication}

The dataset combines transcribed piano performance MIDI files from multiple sources. The same performance could appear multiple times, either transcribed by different models or uploaded originally under different titles. Duplicates do not add new information and distort the performance data distribution. 

For each piece in \textbf{PianoCoRe-C} with multiple performances, the performances are compared pairwise using a content-based heuristic developed to detect and cluster identical or nearly identical performances based on close note onsets. Steps are as follows:
\begin{enumerate}
    \item \textbf{Note Representation:} For each MIDI performance, extract all notes, sort them by time, shift timings so the first note starts at zero, and group notes by pitch number.
    
    \item \textbf{Pairwise Similarity:} Take two performances $x$ and $z$. For each note $x_i$ in $x$ with pitch $p_i = p$, find the closest by onset time matching note $z_j$ in $z$ with the same pitch $p_j = p$. Then, count the number of note pairs whose absolute time difference is below a threshold $\Delta o_{ij} = |o_i(x) - o_j(z)| \le 0.05$ (50 ms, an error bound for a near-perfect onset prediction accuracy in AMT \citep{kong2021high}). The similarity score is the ratio of close note pairs to the total number of notes in $x$. This score was computed in both directions, from $x$ to $z$ and from $z$ to $x$, and the maximum was taken.

    \item \textbf{Clustering:} Performances with at least 50\% similar (close in time) notes are clustered. One ``lead'' file is kept, prioritizing the source datasets with fewer performance samples (GiantMIDI → ATEPP → PERiScoPe → Aria-MIDI) and, when available, alignment recall.
\end{enumerate}

Applying this method flagged 34,452 near-duplicates, which were removed from the \textbf{PianoCoRe-C} dataset, leaving only lead and unique performances. The duplicates are marked in the metadata.

\subsection{MIDI Quality Assessment}
\label{subsec:data-quality}

Besides duplicates, MIDI files transcribed from audio can vary in quality. Since transcription models are trained on limited ground-truth data, they often fail in unseen acoustic conditions \citep{edwards2024data, hu2024towards}. While prior work has proposed perceptually validated metrics \citep{ycart2019peamt, simonetta2022perceptual} and analytical tools \citep{hu2024towards} for evaluating transcriptions, these methods are reference-based and require ground-truth data for comparison.

Heuristics such as NOMML \citep{lee2025gigamidi} have been used to detect inexpressive MIDI data, but they can struggle with transcriptions. In the experiments, NOMML flagged only 29 performances in PianoCoRe as inexpressive. Transcription artifacts, such as onset jitter, create enough  variation to mask a constant tempo, causing score-like performances to appear expressive. 

Not all source MIDI performances in PianoCoRe have corresponding audio or musical scores. To classify each performance, a classifier that assesses MIDI quality directly, independent of score and audio alignment, is trained. The main goal is to detect \textbf{corrupted} transcriptions and \textbf{score-like} performances transcribed from audio synthesizing inexpressive scores.

\subsubsection{Note Alignment and MIDI Quality}
\label{subsubsec:note-alignment-quality}

\begin{figure*}[htb]
    \centering
    \includegraphics[width=\textwidth,alt={Scatter plots mapping performance-to-score note and alignment ratios}]{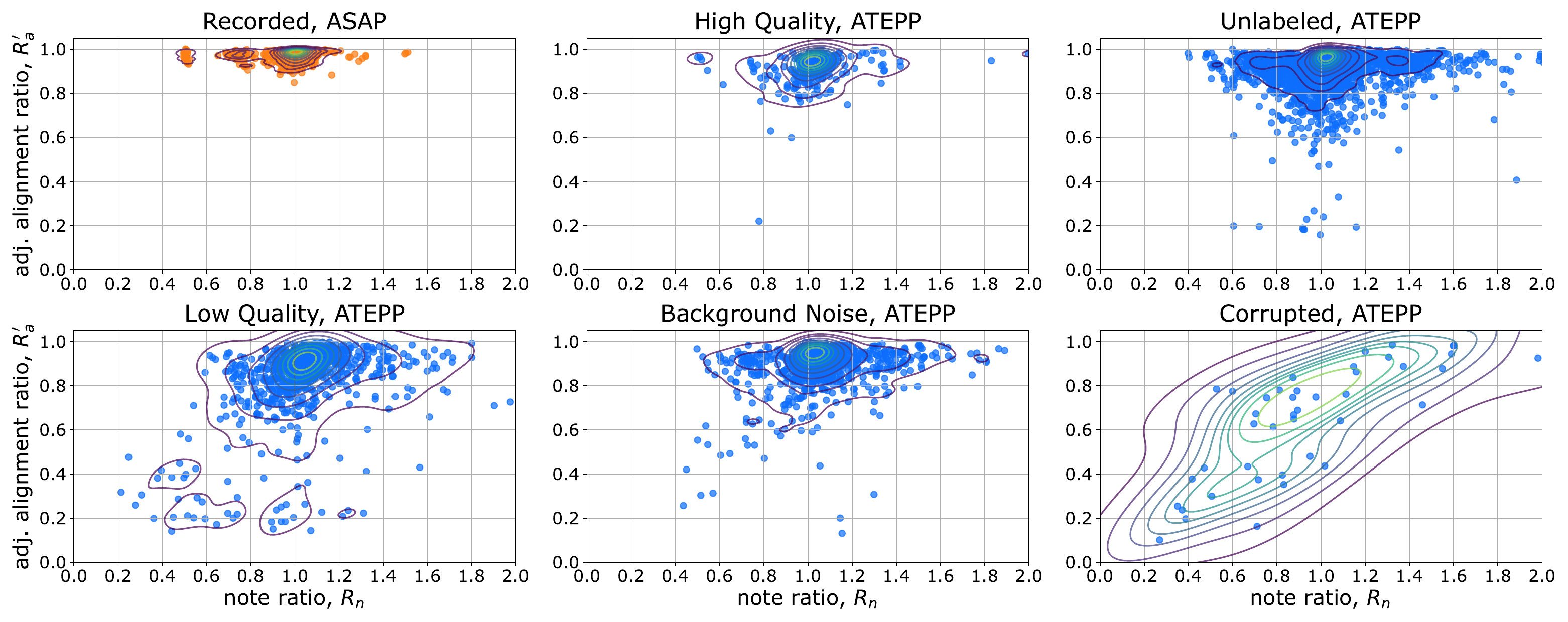}
    \caption{
        MIDI performances from ASAP (orange) and ATEPP (blue) grouped by original labels and mapped as a function of performance-to-score note ratio $R_n$ and adjusted alignment ratio $R_a'$.
    }
    \label{fig:quality-original-labels}
\end{figure*}

The initial hypothesis is that a proxy for MIDI quality is its alignment with the score. The analysis began by examining the differences between recorded performances in ASAP and transcribed performances in ATEPP. In ATEPP, 28.3\% of sequences are labeled as `high quality', `low quality', `background noise', or `corrupted'. Figure~\ref{fig:quality-original-labels} vizualizes the performances using the note ratio $R_n = N_p / N_s$ and adjusted alignment ratio $R_a' = \max(R_a, P_a)$ (Section~\ref{subsec:definitions}). This formulation rewards performances that fully align with the score, even if some segments are not performed.

As we see in Figure~\ref{fig:quality-original-labels}, `recorded' and `high quality' performances cluster in the upper part ($R_a' > 0.85$), indicating strong alignment with the scores. In contrast, `corrupted' files are inconsistently scattered, including both well- and poorly-aligned performances, while `low quality' and `background noise' sequences overlap with high quality and corrupted transcriptions.

Manual inspection of the MIDI files revealed inconsistencies in the original ATEPP labels. Some `low quality' and `unlabeled' files with poor alignment (e.g., \texttt{02709.mid}, \texttt{03001.mid}, \texttt{10193.mid}) contain clearly broken transcriptions. In contrast, a few files labeled as `corrupted' (e.g., \texttt{01591.mid}, \texttt{05389.mid}) align well and are musically usable. Thus, the existing audio-based labels do not reliably reflect MIDI quality.

\subsubsection{MIDI Quality Training Dataset}
\label{subsubsec:midi-quality-dataset}

Based on the analysis of alignments and the adjusted ratio $R_a'$, a soft data labeling heuristic is proposed. Combined with score and recorded MIDI files, the four quality classes are defined as follows:
\begin{enumerate}
    \item \textbf{Score (S):} deadpan score MIDI performances;
    \item \textbf{High Quality (HQ):} any recorded MIDI, transcribed MIDI with $R_a' > 0.9$;
    \item \textbf{Low Quality (LQ):} transcribed, $0.7 < R_a' < 0.85$;
    \item \textbf{Corrupted (C):} transcribed, $R_a' < 0.65$.
\end{enumerate}
The quality ranges are chosen to be disjoint at the boundaries to create clearer distributions for training.

The heuristic was applied to label the deduplicated performances aligned with musical scores. Table~\ref{tab:quality-heuristics} shows the distribution of the soft quality labels.

\begin{table}[htb]
    \centering
    \begin{tabular}{cccc}
        \toprule
        \textbf{HQ} & \textbf{LQ} & \textbf{C} & \textbf{No Label} \\
        \midrule
        170,312 & 4,545 & 140 & 40,597  \\
        \bottomrule
    \end{tabular}
    \caption{
        Distribution of MIDI quality labels computed using the alignment-based heuristics for the deduplicated, aligned performances in PianoCoRe-B.
    }
    \label{tab:quality-heuristics}
\end{table}

These data were used to sample subsets for training, testing and calibration. To ensure composition leakage, a piece-based split was applied, maximizing the number of the real corrupted samples in the test set. Second, to create a diverse dataset, there are no more than three samples for each musical piece from each data source (ASAP, ATEPP, PERiScoPe and Aria-MIDI), as well as a soft quality label (HQ, LQ and C).

As seen in Table~\ref{tab:quality-heuristics}, LQ and C soft labels are underrepresented. For training, 2,500, 1,000 and 86 real HQ, LQ and C samples, respectively, are balanced with synthetic performances built from the sampled HQ MIDI files. The artificial corruptions for LQ/C classes included random note removal (15--25\% / 35--50\%), onset/offset jitter (up to 20 ms / 150 ms), velocity jitter (up to 5 / 20 bins) and random note insertions (up to 5\% / 30\%). Similarly, 953 real scores were augmented with 1,447 synthetic versions (randomized constant velocities, 10 ms onset jitter) to simulate transcription artifacts for score-based audio.

For validation and testing, 200 real Score, HQ and LQ samples are selected alongside 54 Corrupted performances. The classifier calibration set includes all of the real samples from the evaluation split, with no more than three samples per piece, source, and class. The distributions per each set are shown in Table~\ref{tab:midi-quality-dataset}.

\begin{table}[htb]
    \centering
    \begin{tabular}{l cccc}
        \toprule
         & \textbf{S} & \textbf{HQ} & \textbf{LQ} & \textbf{C} \\
        \midrule
        training & 2,500 & 2,500 & 2,500 & 2,500   \\
        \quad real & 953 & 2,500 & 1,000 & 86   \\
        \quad synth & 1,547 & 0 & 1,500 & 2,414   \\
        \midrule
        test & 200 & 200 & 200 & 54  \\
        calibration & 662 & 6,525 & 893 & 54  \\
        \bottomrule
    \end{tabular}
    \caption{
        MIDI quality classification dataset splits.
    }
    \label{tab:midi-quality-dataset}
\end{table}

\subsubsection{MIDI Quality Classifier}
\label{subsubsec:midi-quality-classifier}

The data representation consists of a stacked sequence encoding with five note features: Pitch, TimeShift (s), Velocity (MIDI bins), Duration (s), and absolute TimePosition (s). This encoding does not contain any score features (beat positions and durations) to make the model score-agnostic and universal.

The backbone is a 12-layer transformer encoder \citep{vaswani2017attention} with 80 million parameters, pre-trained using a multi-mask language modeling objective \citep{borovik2025symupe}.
The model dimension is set to 768 and self-attention is extended with Rotary positional embeddings \citep{su2024roformer}. Real-valued note features are passed to sinusoidal embeddings \citep{guo2023fme} for lossless encoding. For classification, penultimate-layer embeddings are prepended with a \texttt{[CLS]} token and processed by a one-layer transformer (dimension 128) and a classification head.

The pre-training was conducted on the \texttt{deduped} subset of Aria-MIDI \citep{bradshaw2025aria} with 371,053 diverse piano MIDI files, provided with the official dataset release. The maximum context length is set to 512 notes. The pre-training included 600,000 steps with batch size 128, while the fine-tuning took 20,000 steps with batch size 512. Training data augmentation included pitch shift ($\pm 6$ semitones), velocity shift ($\pm 6$ MIDI bins) and tempo stretching ($\pm 5\%$).

The MLM backbone was verified on emotion and pianist classification tasks. On the EMOPIA dataset \citep{hung2021emopia}, the classifier achieved a test accuracy of 72.7\% and an F1 score of 72.1\%. On the Pianist8 dataset \citep{chou2024midibert}, the accuracy and F1 score were 86.4\% and 85.5\%. The metrics are close to similarly sized models \citep{liang2024pianobart} and slightly below those of larger models \citep{bradshaw2025scaling}.

\subsubsection{Results}
\label{subsubsec:midi-quality-results}

Table~\ref{tab:midi-quality} shows the evaluation results of the classifier configurations tested on the balanced test set.

\begin{table}[htb]
    \centering
    \resizebox{\columnwidth}{!}{%
    \begin{tabular}{l ccccc}
        \toprule
        \textbf{Model} & \textbf{S} & \textbf{HQ} & \textbf{LQ} & \textbf{C} & \textbf{Avg.} \\
        \midrule
        base & \bfseries 1.000 & \bfseries 0.839 & 0.777 & \bfseries 0.946 & \bfseries 0.891 \\
        \midrule
        no synth & \bfseries 1.000 & 0.759 & \bfseries 0.778 & \bfseries 0.946 & 0.871 \\
        mean & \bfseries 1.000 & 0.828 & 0.752 & 0.881 & 0.865 \\
        mean, no TL & 0.993 & 0.802 & 0.713 & 0.851 & 0.840 \\
        no MLM & 0.995 & 0.773 & 0.667 & 0.842 & 0.819 \\
        \midrule
        mask Pitch & \bfseries 1.000 & 0.803 & 0.723 & 0.913 & 0.860 \\
        mask Timing & 0.990 & 0.788 & 0.747 & 0.851 & 0.844 \\
        mask Velocity & \bfseries 1.000 & 0.834 & 0.776 & 0.893 & 0.876 \\
        \bottomrule
        \end{tabular}
    }
    \caption{
        Evaluation of MIDI quality classifiers using F1 scores. Best scores in \textbf{bold}. no synth -- no synthetic training data, mean -- mean pooling (no \texttt{[CLS]}), no TL -- no transformer layer before the classifier head, no MLM -- token embeddings and classifier only. The last block shows feature-masking ablations.
    }
    \label{tab:midi-quality}
\end{table}

The best configuration achieved a macro F1 score of 89.1\% on the held-out test set. It learned to perfectly distinguish score-like MIDI and showed less errors between HQ, LQ and C classes. The synthetic training samples and token-based aggregation helped to learn more robust decision boundaries. Masking of note features revealed the shared contribution of pitch, dynamic, and timing to MIDI quality classification. Since note-level alignment is imperfect and quality is continuous rather than discrete, errors on the test set are expected.

\subsection{Classifying PianoCoRe-C Dataset}
\label{subsubsec:pianocore-classification}

The best-performing classifier was taken and calibrated on the held-out calibration set (Table~\ref{tab:midi-quality-dataset}). To maximize recall, the sequences are labeled as Corrupted or Score, if the classifier was activated in at least one segment ($p_{\mathrm{S}} > 0.3$ or $p_{\mathrm{C}} > 0.3$). For the LQ class, a conservative threshold of $p_{\mathrm{LQ}} > 0.75$, which does not categorize half of the data as low quality, was chosen. Note that HQ and LQ labels are advisory, as `low quality' MIDI files may be suitable for certain applications. However, the files labeled as Corrupted or Score are in most cases indeed either broken, or were transcribed from rendered scores with constant tempo and/or dynamics. It is better to filter them during piano expression analysis.

The final distribution of MIDI quality labels in the PianoCoRe-C dataset is shown in Table~\ref{tab:data-quality}.

\begin{table}[htb]
    \centering
    \begin{tabular}{l cccc}
        \toprule
        \textbf{Source} & \textbf{S} & \textbf{HQ} & \textbf{LQ} & \textbf{C} \\
        \midrule
        ASAP & 0 & 1,066 & 0 & 0 \\
        ATEPP & 0 & 10,231 & 900 & 433 \\
        GiantMIDI & 11 & 2,071 & 52 & 5 \\
        PERiScoPe & 82 & 34,596 & 91 & 4 \\
        Aria-MIDI & 1,151 & 180,977 & 18,359 & 17 \\
        \midrule
        PianoCoRe & 1,244 & 228,941 & 19,402 & 459 \\
        \bottomrule
    \end{tabular}
    \caption{
        PianoCoRe dataset and its source subsets labeled by the MIDI quality classifier.
    }
    \label{tab:data-quality}
\end{table}

\subsection{PianoCoRe-B Dataset}
\label{subsec:pianocore-b}

By applying the deduplication and quality assessment models to \textbf{PianoCoRe-C} dataset, we obtain \textbf{PianoCoRe-B}. The filtered subset consists of 214,092 deduplicated performance MIDI not classified as Corrupted or Score. There are 5,591 musical pieces composed by 478 composers (Table~\ref{tab:dataset-comparison}). 

\subsubsection{Applications}
\label{subsubsec:pianocore-b-applications}

\textbf{PianoCoRe-B} is designed for tasks that depend on large amounts of clean and reliable piano performance data. Specifically, this dataset is useful for large-scale, self-supervised pre-training, musical analysis of performance styles, and piano performance generation.

\section{Refined Note Alignment}
\label{sec:refined-alignment}

Piano expression modeling tasks require precise note-level alignment between scores and performances. The \textbf{PianoCoRe-A/A*} subsets (`A' for `Aligned') consist of all performance MIDI files that are temporally aligned to scores. Two forms of alignment are considered:
\begin{enumerate}
    \item \textbf{Raw Alignments:} processed output of Parangonar, containing matches, insertions, and deletions between score and performance notes;
    
    \item \textbf{Refined Alignments:} raw alignments, refined using the \textbf{RAScoP} pipeline, which cleans and completes initial matches.
\end{enumerate}

\subsection{Raw Note Alignment Challenges}
\label{subsec:alignment-challenges}

\ifpdf
\begin{figure*}[!htb]
    \centering
    \includegraphics[width=\textwidth,alt={Visualizations of note-level alignment errors and missing score segments}]{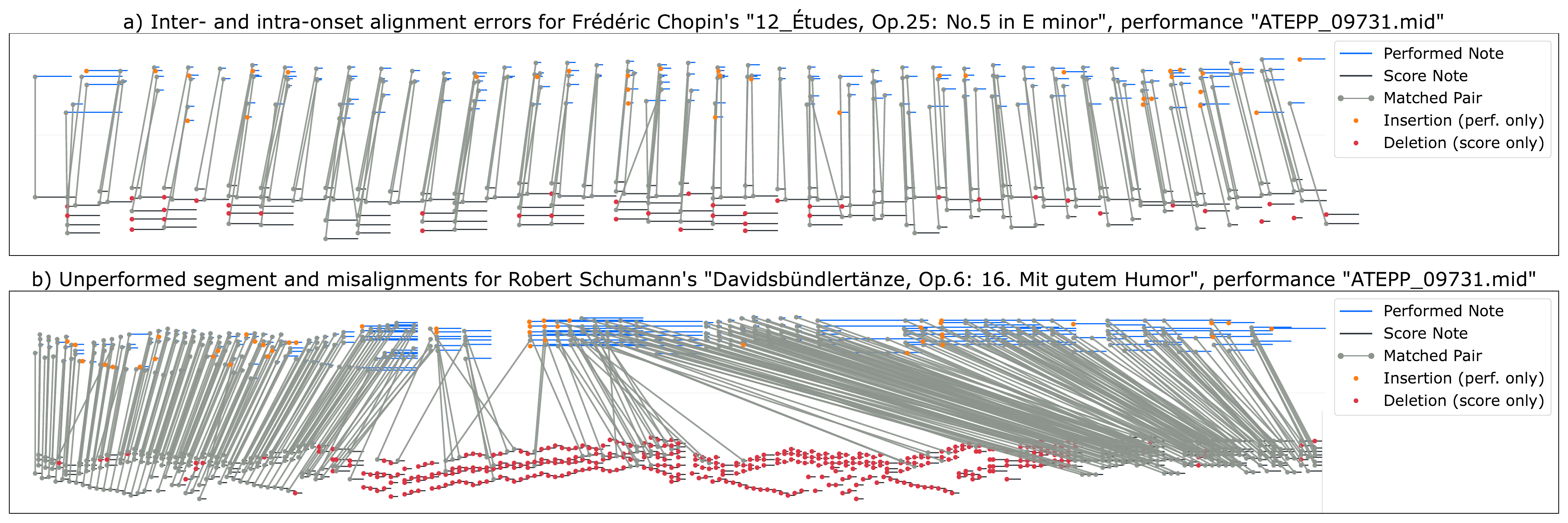}
    \caption{
        Real-world alignment challenges motivating the RAScoP pipeline. Top: local timing errors (crossed links) and missing/extra notes. Bottom: large structural deviation from a missing score segment, causing incorrect links.  Other performed notes remain usable. Alignments were computed with Parangonar.
    }
    \label{fig:alignment-errors}
\end{figure*}
\else\fi

A direct output from note aligners like Parangonar \citep{peter2023parangonar} or Nakamura's alignment tool \citep{nakamura2017alignment}, while powerful, is sometimes insufficient for direct use in generative models. Raw alignments can suffer from issues, illustrated in Figure~\ref{fig:alignment-errors}:

\begin{itemize}
    \item \textbf{Temporal Discontinuities:} Incorrect alignment links that cross in time or match musically distant notes, leading to unrealistic tempo fluctuations and high inter-onset timing deviations;
    
    \item \textbf{Alignment Holes:} Continuous regions of unaligned notes in the score or performance, often caused by skipped repeats or transcription errors.
\end{itemize}

\ifpdf\else
\begin{figure*}[!htb]
    \centering
    \includegraphics[width=\textwidth,alt={Visualizations of note-level alignment errors and missing score segments}]{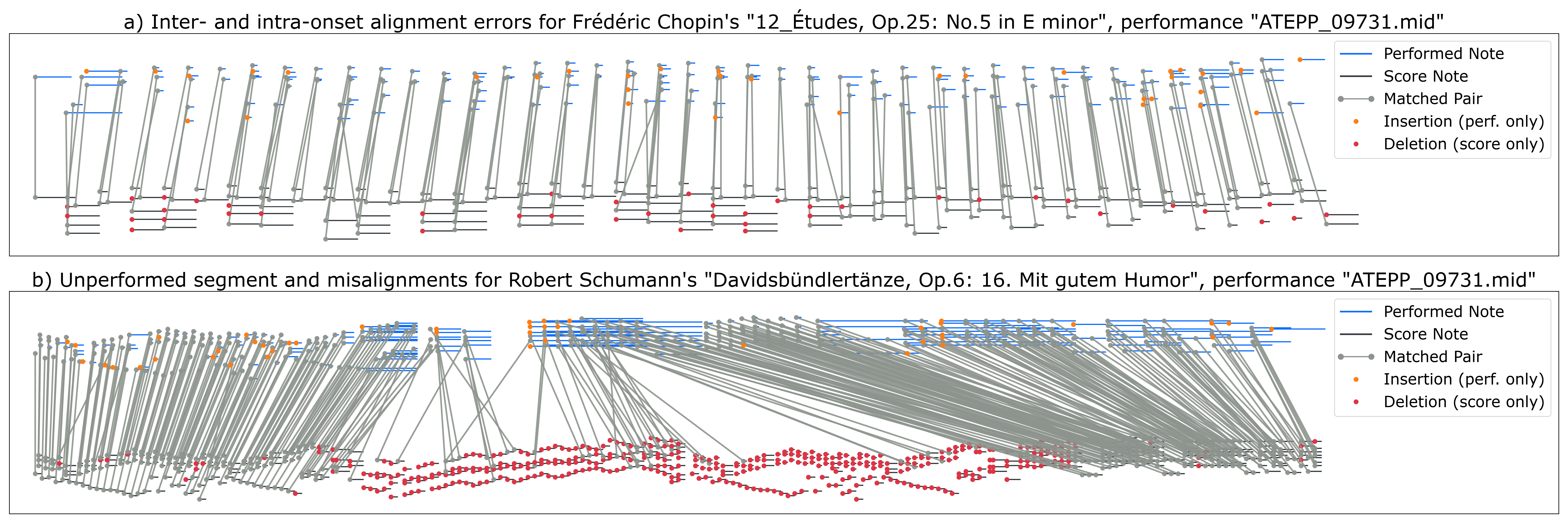}
    \caption{
        Real-world alignment challenges motivating the RAScoP pipeline. Top: local timing errors (crossed links) and missing/extra notes. Bottom: large structural deviation from a missing score segment, causing incorrect links.  Other performed notes remain usable. Alignments were computed with Parangonar.
    }
    \label{fig:alignment-errors}
\end{figure*}
\fi

Some performance rendering models were trained only on a subset of aligned score and performance notes with incomplete score contexts \citep{rhyu2022sketching, zhang2024dexter, tang2025towards}. Other models removed timing outliers \citep{xia2016expressive, jeong2019virtuosonet} and interpolated missing notes \citep{borovik2023scoreperformer, borovik2025symupe}. However, these processes are not available as easy-to-use tools.

A configurable algorithm was designed to create a parallel score and performance dataset by cleaning evident outliers and interpolating notes for which no performance counterpart exists. Specifically, this algorithm addresses two main problems: 
\begin{itemize}
    \item \textbf{Timing Errors:} remove large inter- and intra-onset deviations and implied unrealistic tempi;
    
    \item \textbf{Missing Notes:} fill in the unperformed notes to have complete performed score contexts.
\end{itemize}

The following section describes this algorithm.

\subsection{Alignment Cleaning and Refinement}
\label{subsec:alignment-refinement}

\begin{figure*}[htb]
	\centering
	\includegraphics[width=0.97\textwidth,alt={Flowchart illustrating the four stages of alignment refinement}]{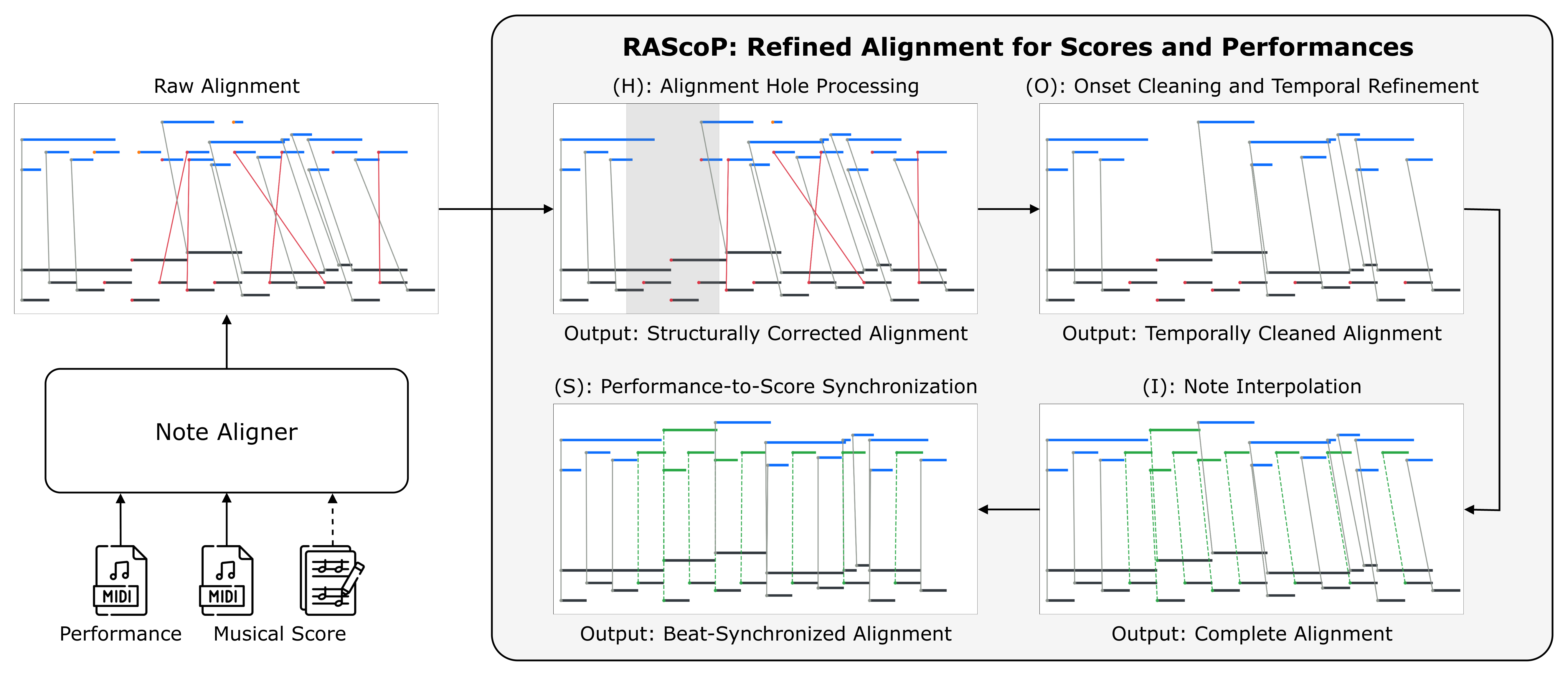}
	\caption{
        Note-level alignment and the RAScoP pipeline for alignment refinement. The processing steps are demonstrated using an artificial example containing all types of errors. Score notes are drawn in black and performance notes are drawn in blue and green.
    }
	\label{fig:rascop}
\end{figure*}

\textbf{RAScoP} (Refined Alignment for Scores and Performances) is an integrated pipeline designed to take a raw score-performance alignment and transform it into a clean, complete, and temporally coherent parallel score-performance data pair. The algorithm analyzes and refines the alignment through four sequential steps, illustrated in Figure~\ref{fig:rascop}:

\begin{enumerate}
    \item \textbf{(\texttt{H}):} alignment hole processing;
    \item \textbf{(\texttt{O})}: onset cleaning and temporal refinement;
    \item \textbf{(\texttt{I}):} note interpolation;
    \item \textbf{(\texttt{S}):} performance-to-score synchronization.
\end{enumerate}

\subsubsection{Alignment Hole Processing}
\label{subsubsec:hole-processing}

The first step detects and removes large, structurally incorrect alignment sections. An `alignment hole' is defined as a continuous region of notes where the alignment is sparse or nonsensical (only a few notes are aligned). In scores, the holes correspond to unperformed score measures (e.g., repeats), whose individual notes may be incorrectly matched with random performance notes. In the performances, the holes are the extra performed segments whose notes may be inadvertently aligned with random score notes.

To detect holes, a sliding window approach is used. Let $H_a$ be a ratio of unaligned notes within a surrounding window of $H_w$ notes for a given note. If $H_a$ ratio exceeds a threshold $H_r$, the note is flagged. Contiguous regions of flagged notes are designated as holes, and all alignment pairs within them are removed.

The default values are $H_w = 31$ notes and $H_r = 0.75$. The window size is close to double the median (15) and mean (16.9) number of notes in a measure in all scores in the dataset. With this window, we consider on average one measure to the left and one to the right. Setting the threshold at $75\%$ ensures that only regions that are almost entirely unaligned are removed. 

\subsubsection{Onset Cleaning and Temporal Refinement}
\label{subsubsec:onset-cleaning}

This stage refines the temporal alignment of concurrently played notes (chords) and corrects large-scale time shifts. First, all aligned notes are used to build the initial onset pair list: tuples of score onset beat $o_i$ and the average performed onset time $t(o_i)$ for all notes in the chord. Then, note and onset times are checked for misalignments and outliers based on: 
\begin{itemize}
    \item high intra-onset deviations;
    \item inter-onset intervals that deviate from the local performance tempo. 
\end{itemize}

For intra-onset deviations, the onset deviations $\Delta t_i(n_j) = t(n_j) - t(o_i)$ from $t(o_i)$ are computed for all notes $n_j$ in a chord: $\{n_j | o(n_j) = o_i \}$. By default, notes whose onsets deviate from $t(o_i)$ by more than two standard deviations are removed from the alignment as outliers. For chords with two distant notes, both notes will be removed if the condition is met. 

For inter-onset intervals, the method estimates the maximum and minimum plausible time shifts $\Delta t_{\max}(o_i)$ and $\Delta t_{\min}(o_i)$ between the current and previous score onsets $o_i$ and $o_{i-1}$. If the time interval between the current and previous onset implies a tempo outside a plausible range (by default, 15--480 BPM), it is identified as an alignment jump. This onset can be filtered out of the alignment. However, by default, the timing of the notes of the affected onsets is adjusted.

First, a local tempo $\tau_{\mathrm{local}}(o_i)$ is estimated based on a $w$-second window (by default, $w = 8$) of preceding performed note onsets $\mathrm{O}_{\mathrm{local}}(o_i) = \{o_j | t(o_i) - t(o_j) < w\}$. Then, the expected onset time $\hat{t}(o_i)$ is computed using the inter-onset beat shift $\mathrm{IOI}_i^s = o_i - o_{i-1}$ and $\tau_{\mathrm{local}}(o_i)$. Using the expected onset time, the required time shift $\Delta t_{\mathrm{adj}}(o_i) = \hat{t}(o_i) - t(o_i)$ is determined, and the subsequent performance notes are shifted accordingly.

This step explicitly alters the global timing in the original performance MIDI. However, after the shift, the onsets fall into the range of the plausible local tempos, and tempo outliers are not learned by the trained models. Any unperformed notes can be also naturally filled in with the same local performance tempo.

In addition, close onset pairs with $\Delta t(o_{i}) = t(o_i) - t(o_{i-1}) < 0.01$ (10 ms) are filtered out to avoid two same-pitch note-on events (which is impossible for a human performer). After the alignment hole processing and onset cleaning, the algorithm cleans up the performance MIDI by removing notes without a link in the alignment. In the end, only matched and cleanly performed notes remain. 

\subsubsection{Note Interpolation}
\label{subsubsec:note-interpolation}

This step interpolates the unperformed notes to create parallel note-aligned score-performance pairs.

The note onset time $t(n_i)$ of a note $n_i$ is linearly interpolated from two neighboring performed notes $n_j$ and $n_k$. To avoid the contribution of very close notes, the configurable minimum beat and time intervals $n_j$ and $n_k$ between the two anchor notes are used ($t(n_k) - t(n_j) \ge \Delta t_{\mathrm{int}}$ and $o(n_k) - o(n_j) \ge \Delta o_{\mathrm{int}}$).

Note articulation (duration) and dynamics (MIDI velocity) are averaged and weighted by the performed notes in the neighboring beats. The weights are inversely proportional to the absolute beat distances $\mathrm{IOI}_{i,j}^{s} = |o(n_j) - o(n_i)|$ from the score position of the note $n_i$ being interpolated. Closer notes contribute a higher weight to the interpolated features.

The algorithm prevents the creation of notes with identical pitch and onset, and shortens overlapping notes so that at each new key press the previous note is closed. The result is a performance MIDI file aligned with the score at the note level. Interpolated notes are marked with a special MIDI text marker, allowing them to be filtered out or marked during model training. 

\subsubsection{Performance-Score Synchronization}
\label{subsubsec:synchronization}

This step synchronizes the beat structure of the refined performance MIDI with the score. This data format is commonly used in MIDI encodings with beat/bar tempo \citep{huang2020pop, hsiao2021compound, zeng2021musicbert}. The alignment pairs are used to compute a beat-to-time mapping and insert inter-beat tempo changes into the performance MIDI. For example, for a 4/4 time signature and 480 ticks per quarter, notes at the beats are separated by 480 ticks in both the score and performance MIDI, with exact times derived from tempo changes.

Finally, the entire performance is shifted so that its first played note occurs at the same time as the first score note, ensuring a consistent starting point for all performances of the same composition.

\subsubsection{Final Output}
\label{subsubsec:refinement-output}

The algorithm returns the refined alignment, refined performance MIDI and note-level alignment recall ratios. The recall values from different stages (initial, hole processing, and onset cleaning) serve as quantitative indicators of alignment quality and can be used to interrupt the refinement process. The alignment is released as a compressed \texttt{.npz} file containing an array of performance note indices aligned to the sorted score MIDI notes, along with a boolean mask for interpolated notes. 
All MIDI processing steps are performed using the \texttt{symusic} Python library \citep{liao2024symusic}.

The presented refinement does not rematch links produced by the note aligner. It only filters existing links and interpolates missing notes. Each step of the pipeline can be enabled independently. Default parameters were chosen empirically rather than optimized, as automated evaluation would require precise human annotations. Custom clean datasets can be generated using the released raw alignments and MIDI files.

In PianoCoRe, all refined performance MIDI files underwent the first three stages of the RAScoP alignment and the final initial performance onset shift. Beat synchronization was not applied in order to preserve the original timing without re-quantizing the note onsets and offsets. Synchronization can be computed using the refined score MIDI, performance MIDI and note-level alignment.

\subsection{Refinement Quality Evaluation}
\label{subsec:refinement-evaluation}

To quantitatively demonstrate the effectiveness of \textbf{RAScoP}, the trade-off between alignment temporal integrity (the distribution of intra- and inter-onset deviations) and alignment recall ($R_a$) is evaluated. 

\begin{figure}[htb]
	\centering
    \includegraphics[width=1.\columnwidth,alt={Box plots showing inter-onset deviations and beat tempos}]{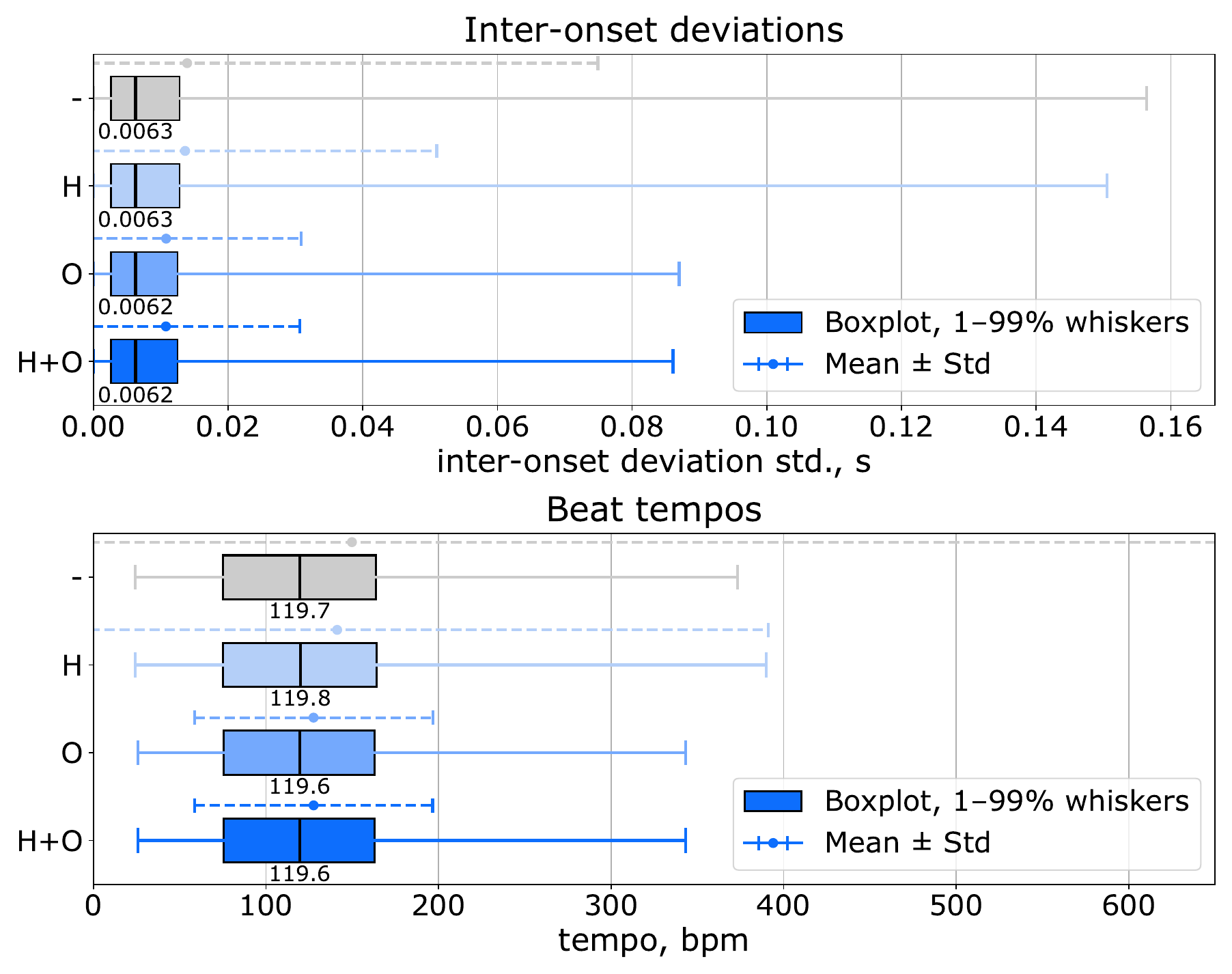}
    \caption{
        Distribution of inter-onset deviations and beat tempos for alignments before processing (-), after hole processing (H), after onset cleaning (O), and after both hole and onset cleaning (H+O).
    }
	\label{fig:eval-rascop}
\end{figure}

The benefit of alignment refinement is shown in Figure~\ref{fig:eval-rascop}. Applying the full pipeline (H+O) significantly reduces the standard deviation of inter-onset deviations within chords, indicating cleaner note timing patterns. Furthermore, the distribution of beat tempos becomes more stable and centered around a musically plausible range, as the algorithm corrects for the extreme tempo values implied by raw, noisy alignments.

Table~\ref{tab:eval-match-ratio} quantifies the `cost' of the cleaning process in terms of alignment recall. The performances are grouped from higher to lower recall. Overall, the average recall $\overline{R}_a$ decreases by a modest 1.5\% (from 0.935 to 0.920), with the Onset Cleaning stage (O) contributing most to this reduction. The cleaning process primarily affects the highest-quality alignments ($\overline{R}_a>0.95$), reducing their share from 54.3\% to 42.9\%. These sequences are not discarded, but rather migrate to the still-high-quality lower bands. After refinement, the majority of sequences (86.6\%) still maintain a high alignment recall of over 85\%. The loss of a few alignment links is an acceptable price for the improvement in the temporal quality of the performance data.

\begin{table}[htb]
    \small
    \centering
    \resizebox{\columnwidth}{!}{
    \begin{tabular}{c cc cc cc}
        \toprule
        & \multicolumn{2}{c}{Raw} & \multicolumn{2}{c}{After H} & \multicolumn{2}{c}{After H+O} \\
        \cmidrule(lr{0.125em}){2-3} \cmidrule(lr{0.125em}){4-5}
        \cmidrule(lr{0.125em}){6-7}
        Band & $\overline{R}_\mathrm{A}$ & \% & $\overline{R}_\mathrm{H}$ & \% & $\overline{R}_{\mathrm{H+O}}$ & \% \\
        \midrule
        0.95--1.00 & 0.975 & 54.3 & 0.975 & 53.9 & 0.973 & 42.9 \\
        0.90--0.95 & 0.929 & 26.6 & 0.929 & 26.7 & 0.928 & 30.4 \\
        0.85--0.90 & 0.879 & 10.1 & 0.878 & 10.0 & 0.878 & 13.3 \\
        \midrule
        0.80--0.85 & 0.828 & 4.7 & 0.828 & 4.6 & 0.828 & 6.5 \\
        0.75--0.80 & 0.779 & 2.1 & 0.778 & 2.2 & 0.777 & 3.2 \\
        0.70--0.75 & 0.725 & 1.1 & 0.727 & 1.0 & 0.728 & 1.6 \\
        \midrule
        0.60--0.70 & 0.660 & 0.7 & 0.663 & 1.1 & 0.661 & 1.5 \\
        0.00--0.60 & 0.471 & 0.4 & 0.464 & 0.5 & 0.462 & 0.6 \\
        \midrule
        all & 0.935 & 100.0 & 0.934 & 100.0 & 0.920 & 100.0 \\
        \bottomrule
    \end{tabular}
    }
    \caption{
        Mean alignment recall $\overline{R}$ after different alignment refinement stages and the ratio of sequences (\%) inside different recall bands.
    }
    \label{tab:eval-match-ratio}
\end{table}

\subsection{PianoCoRe-A Dataset}
\label{subsec:pianocore-a}

Applying this pipeline to the files from \textbf{PianoCoRe-B} yields the final aligned datasets. \textbf{PianoCoRe-A} contains 157,207 cleaned and note-aligned sequences from \textbf{PianoCoRe-B} for 1,591 pieces written by 151 composers, totaling 12,509 h of music (Table~\ref{tab:dataset-comparison}). 

The performances can be filtered out for any applications based on the alignment ratio. For tasks that demand the highest possible data fidelity, \textbf{PianoCoRe-A*} is introduced. This is a high-confidence subset of \textbf{PianoCoRe-A} containing High Quality MIDI with at least 85\% of aligned notes. \textbf{PianoCoRe-A*} consists of 130,275 performances for 1,517 pieces.

\subsubsection{Applications}
\label{subsubsec:pianocore-a-applications}

\textbf{PianoCoRe-A/A*} represent a large-scale resource of score-performance-aligned piano MIDI data. They pave the way for training more nuanced models for rendering expressive piano performances without having to perform rigorous data matching and alignment.

\section{Music Performance Rendering}
\label{sec:music-performance}

The PianoCoRe dataset is validated on a downstream task of expressive piano performance rendering. The hypothesis is that the scale, diversity, and targeted refinement of the \textbf{PianoCoRe-A} dataset enable the training of more accurate performance models compared to baselines trained on smaller or uncleaned data subsets.

\subsection{Experimental Setup}
\label{subsec:experimental-setup}

The experiments used PianoFlow \citep{borovik2025symupe}, a model for symbolic music performance rendering based on conditional flow matching \citep{lipman2022flow}. It employs an encoder transformer to inpaint masked performance features $x_{\text{m}}$ (TimeShift, Velocity, TimeDuration, TimeDurationSustain) given score features $y$ (Pitch, Position, PositionShift, and Duration) and performance context $x_{\text{ctx}}$. As Aria-MIDI does not distinguish between pressed and sustained notes, only seven features without TimeDuration were used. 
The base configuration (8 layers, 24 million parameters) was adopted, and a learned embedding was added to interpolated notes, as in the original model.

The model was trained on subsets of aligned and cleaned performances from PianoCoRe-A: ASAP, ASAP+ATEPP, ASAP+ATEPP+PERiScoPe, and the full dataset. Performances with fewer than 85\% aligned notes ($R_{\textrm{RAScoP}} < 0.85$) were removed to retain more real played notes. For ablation, models were trained on all PianoCoRe-A performances ($R_{\textrm{RAScoP}} \ge 0.7$) and a version of the dataset without the hole and onset cleaning from RAScoP pipeline (raw alignments plus note interpolation). Data were split by composition into 90\%/10\% for training/evaluation, all movements and performances of a piece appeared in only one split. 

\subsection{Results}
\label{subsec:experimental-results}

\subsubsection{Training Convergence}
\label{subsubsec:training-convergence}

\begin{figure*}[htb]
    \centering
    \includegraphics[width=1.\textwidth,alt={Line graphs showing validation loss curves during model training}]{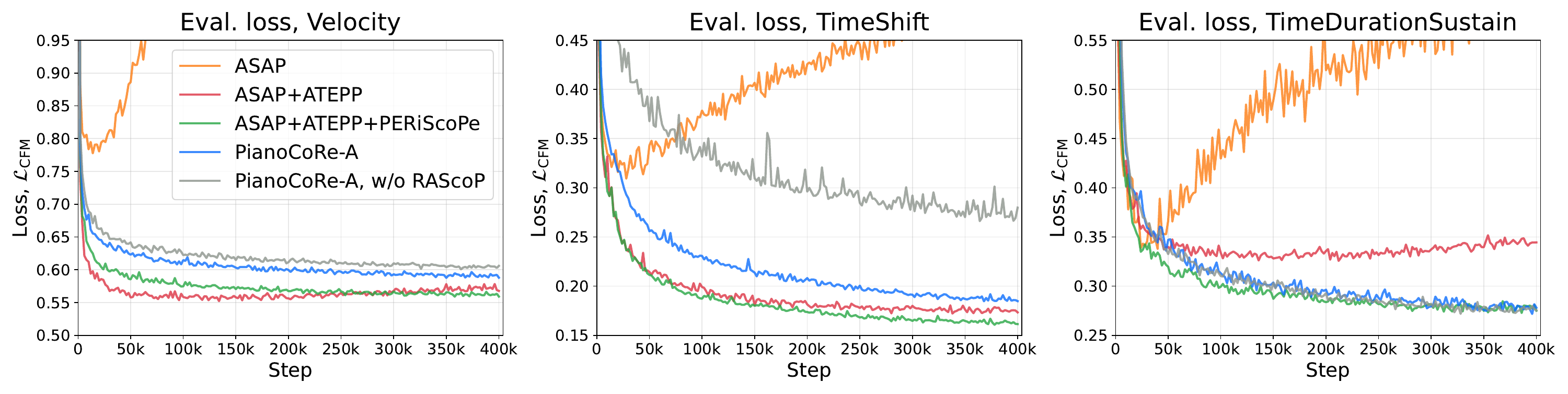}
    \caption{
        Validation loss curves for PianoFlow trained on different subsets of the data. Larger and refined training datasets reduce overfitting in the long run.
    }
    \label{fig:eval-loss}
\end{figure*}

Figure~\ref{fig:eval-loss} illustrates the feature-based validation losses tracked during training. Each model was evaluated on a validation set drawn from the same source data (e.g., the `ASAP+ATEPP' model on unseen `ASAP+ATEPP' performances). The results reveal a pattern: the model trained only on `ASAP' quickly overfits, demonstrating that a small dataset, even of high quality, is insufficient. As the scale of the data increases (`+ATEPP', `+PERiScoPe'), overfitting is delayed.

The comparison between the `PianoCoRe-A' model (blue) and its unrefined counterpart `w/o RAScoP' (gray) provides direct evidence of the value of the refinement pipeline. The refined dataset yields a more stable and consistently lower validation loss, particularly for the note time shifts. This confirms that targeted removal of temporal noise is crucial for learning an accurate timing model. 

\subsubsection{Unconditional Generation}
\label{subsubsec:unconditional-generation}

This section presents the evaluation results for the unconditional performance rendering. The inference set included test set scores with at least three performances from two different MIDI sources (e.g., ASAP and Aria-MIDI). The models rendered each score in its entirety seven times. Pearson correlation \citep{jeong2019graph, borovik2023scoreperformer, zhang2024dexter} between the note features of the dataset and rendered performances was computed. The evaluated features are: onset velocity (Vel), relative inter-onset intervals (IOI), relative intra-onset deviations (OD), and note articulation (Art).

Table~\ref{tab:eval-correlation} presents the mean Pearson correlation between the model outputs and the ground-truth performances from a multi-source test set. Models trained on more diverse datasets (`+ ATEPP`, `+ PERiScoPe`, and `PianoCoRe-A`) consistently outperform the baseline trained only on `ASAP`. Interestingly, the model trained on ASAP and ATEPP shows higher correlation with an average set of performances from PianoCoRe-A. This may be because ATEPP specifically focuses on the performances of renowned pianists. Other datasets contain a wider variety of performance styles.

\begin{table}[htb]
    \newcommand{\valueErr}[2]{%
      #1{\scriptsize $\pm$#2}
    }
    \centering
    \resizebox{\columnwidth}{!}{
    \begin{tabular}{l cccc}
        \toprule
         & \textbf{Vel} & \textbf{IOI} & \textbf{OD} & \textbf{Art} \\
        \midrule
        Dataset & \valueErr{0.57}{0.19} & \valueErr{0.90}{0.06} & \valueErr{0.22}{0.17} & \valueErr{0.44}{0.19} \\
        \midrule
        ASAP & \valueErr{0.37}{0.17} & \valueErr{0.83}{0.11} & \valueErr{0.07}{0.15} & \valueErr{0.28}{0.13} \\
        + ATEPP & \textbf{\valueErr{0.42}{0.16}} & \valueErr{0.85}{0.11} & \textbf{\valueErr{0.12}{0.14}} & \valueErr{0.35}{0.15} \\
        + PERiScoPe & \valueErr{0.41}{0.17} & \textbf{\valueErr{0.86}{0.11}} & \valueErr{0.11}{0.17} & \textbf{\valueErr{0.36}{0.17}} \\
        \midrule
        \textbf{PianoCoRe-A} & \valueErr{0.40}{0.17} & \textbf{\valueErr{0.86}{0.11}} & \valueErr{0.10}{0.17} & \valueErr{0.35}{0.17} \\
        \quad $R_{\textrm{RAScoP}} \ge 0.7$ & \valueErr{0.39}{0.16} & \valueErr{0.85}{0.11} & \valueErr{0.09}{0.16} & \valueErr{0.35}{0.18} \\
        \quad w/o RAScoP & \valueErr{0.41}{0.16} & \valueErr{0.85}{0.11} & \valueErr{0.09}{0.16} & \textbf{\valueErr{0.36}{0.18}} \\
        \bottomrule
    \end{tabular}
    }
    \caption{
        Correlation between the features of the rendered and PianoCoRe-A performances. First row - intra-set correlations, other rows - models trained on different data subsets. \texttt{Vel} - velocity, \texttt{IOI} - inter-onset-interval, \texttt{OD} - relative onset deviation, \texttt{Art} - sustained articulation. The best scores are in \textbf{bold}.
    }
    \label{tab:eval-correlation}
\end{table}

More training data with more interpolated notes ($R_{\textrm{RAScoP}} \ge 0.7$) slightly hurts the unconditional rendering capabilities. The model trained on raw data without the cleanup shows lower correlation with higher quality performances for note timing (IOI and OD). 

\subsubsection{Performance Continuation}
\label{subsubsec:performance-continuation}

The final analysis evaluated the models in a performance continuation task across four distinct test domains: ASAP, ATEPP, PERiScoPe, and Aria. As in the previous experiments, compositions and performances were not seen during the training. The models performed 256 notes in parallel using the performance context of the preceding 256 notes. Table~\ref{tab:eval-conditional} shows the mean absolute error computed against the ground truth performance features.

\begin{table*}[htb]
    \centering
    \resizebox{\textwidth}{!}{
    \begin{tabular}{l c ccc ccc ccc ccc}
        \toprule
        && \multicolumn{3}{c}{\textbf{ASAP}} & \multicolumn{3}{c}{\textbf{ATEPP}} & \multicolumn{3}{c}{\textbf{PERiScoPe}} & \multicolumn{3}{c}{\textbf{Aria-MIDI}} \\
        \cmidrule(lr){3-5} \cmidrule(lr){6-8} \cmidrule(lr){9-11} \cmidrule(lr){12-14}
        \textbf{Dataset} & \textbf{Size} & \textbf{Vel} & \textbf{TS} & \textbf{TD} & \textbf{Vel} & \textbf{TS} & \textbf{TD} & \textbf{Vel} & \textbf{TS} & \textbf{TD} & \textbf{Vel} & \textbf{TS} & \textbf{TD} \\
        \midrule
        ASAP & 1k & 9.885 & 0.023 & 0.187 & 9.928 & 0.022 & 0.206 & 9.893 & 0.023 & 0.230 & 9.957 & 0.027 & 0.275 \\
        + ATEPP & 6k & 9.157 & 0.017 & 0.168 & 8.230 & 0.015 & 0.191 & 8.782 & 0.016 & 0.216 & 8.721 & 0.019 & 0.252 \\
        + PERiScoPe & 25k & 8.851 & \textbf{0.016} & \textbf{0.154} & \textbf{7.888} & \textbf{0.013} & \textbf{0.189} & 8.117 & \textbf{0.015} & \textbf{0.192} & 8.133 & \textbf{0.017} & 0.230 \\
        \midrule
        \textbf{PianoCoRe-A} & 124k & \textbf{8.613} & \textbf{0.016} & 0.155 & 7.967 & 0.014 & 0.194 & 8.094 & \textbf{0.015} & 0.194 & \textbf{7.872} & \textbf{0.017} & \textbf{0.205} \\
        \quad $R_{\textrm{RAScoP}} \ge 0.7$ & 141k & 8.631 & \textbf{0.016} & 0.158 & 7.944 & 0.014 & 0.196 & \textbf{8.071} & \textbf{0.015} & 0.194 & 7.921 & \textbf{0.017} & 0.206 \\
        \quad w/o RAScoP & 124k & 8.734 & 0.017 & 0.159 & 8.059 & 0.015 & 0.193 & 8.199 & 0.016 & 0.196 & 8.055 & 0.018 & 0.211 \\
        \bottomrule
    \end{tabular}
    }
    \caption{
        Conditional performance rendering (performance continuation) results across training subsets and unseen source sequences. \texttt{Size} denotes the training set size. \texttt{Vel} -- Velocity (MIDI bins), \texttt{TS}--TimeShift (s), \texttt{TD} -- TimeDurationSustain (s). Lower is better, best values are in \textbf{bold}.
    }
    \label{tab:eval-conditional}
\end{table*}

The results complement the previous findings. With more training data, the model performs better on MIDI files of different sources. PianoCoRe-A achieves the best average performance on ASAP and Aria-MIDI performances and second-best results on the other subsets. Only the model trained on data without overrepresented Aria-MIDI achieves similar or lower errors on ATEPP and PERiScoPe. Given the validation loss plots in Figure~\ref{fig:eval-loss}, the full dataset model has room for an improvement in the long run. Overall, the results show the potential of PianoCoRe for training performance models robust to varying piano data distributions.

\subsection{Future Work}
\label{subsec:future-work}

A subjective listening test of popular models trained on the subsets of PianoCoRe dataset would be a valuable next step to confirm that objective improvements translate to human perception. Since performances from Aria-MIDI dominate PianoCoRe, a more balanced sampling of performances per source might provide a better generalization to all source data domains. Fine-tuning on high-fidelity subsets, such as ASAP, could potentially improve performance even further.

\section{Limitations}
\label{sec:limitations}

Despite rigorous curation, PianoCoRe has limitations. There are no duplicate musical pieces with different names. However, an error margin of 1\% is reserved for potential movement-level naming errors that were inherited from the source datasets. Furthermore, the dataset distribution remains skewed toward Western classical repertoire and popular composers, reflecting the biases of the underlying open-source corpora.

The dataset relies on open-source MusicXML scores and automated alignment. MusicXML scores are not error-free and may also include a segment of a complete written musical composition. Since it is difficult to validate large-scale datasets precisely, any errors in the source notations may propagate to the downstream applications. Also, due to the iterative combination of source datasets, fewer than 1\% of performances may contain neighboring movements or differ from the scores by more than twice the length. It is recommended to use composition-wise splits in the applications using the dataset.

The classifier-based MIDI quality labels were calibrated for recall in the corrupted and score-like classes to filter out incorrect and inexpressive data. The labels do not guarantee perfect alignment with human expectations. During note interpolation, RAScoP may introduce deadpan performance note segments that must be addressed by downstream applications. Additionally, interpolation does not handle sustain pedal effects. A better solution would be to predict missing notes and pedals using a trained model.

\section{Conclusion}
\label{sec:conclusion}

This article presented \textbf{PianoCoRe}, a unified, large-scale piano MIDI dataset created by combining, refining, annotating, and aligning existing open-source corpora. Released in tiered subsets, PianoCoRe supports a wide spectrum of tasks: from performance analysis and large-scale pre-training to expressive piano performance rendering and score-to-performance translation. The dataset enables reproducible research by allowing researchers to create non-overlapping data splits across previously isolated datasets.

To ensure data integrity, two challenges were addressed: the quality of performance MIDI and note-level alignments. A classifier was trained to identify deadpan and corrupted MIDI transcriptions, and an alignment refinement pipeline was designed to remove temporal outliers in aligned score-performance data. The experiments showed that the model trained on these refined subsets benefits from the increased repertoire diversity and cleaner note features.

Future directions include extending the methodology to multi-instrument repertoires, developing more robust quality assessment models and incorporating more granular score and performance annotations. By making PianoCoRe openly available, the goal is to establish a foundation for advancing symbolic music performance modeling and analysis research.

\section*{Ethical Statement}
\label{sec:ethical-statement}

The curation of large-scale symbolic datasets presents challenges regarding copyright and intellectual property. A best-effort attempt was made to filter PianoCoRe according to European Union public-domain regulations (works whose authors have been deceased for more than 70 years). However, achieving 100\% accuracy across thousands of files from diverse sources is inherently difficult. For transparency, the annotated composer metadata is released alongside the dataset.

The dataset, original and processed files, metadata, and alignment annotations are published under a CC-BY-NC-SA 4.0 license. The license respects the licenses used for the source datasets. No formal ethics approval or human participant consent was required for this study, as it involved the processing of publicly available MIDI data and did not involve human subjects.

\section*{Data Accessibility}
\label{sec:data-accessibility}

The PianoCoRe dataset and related resources are released to ensure reproducibility:
\begin{itemize}
    \item \textbf{Code:} Documentation and usage examples are available at the project repository: \url{https://github.com/ilya16/PianoCoRe}. The source code for the RAScoP pipeline and the MIDI quality classifier is integrated into the \texttt{symupe} library: \url{https://github.com/ilya16/SyMuPe}.
    \item \textbf{Dataset:} The dataset is archived on Zenodo (\url{https://doi.org/10.5281/zenodo.19186016}) and is available on Hugging Face (\url{https://huggingface.co/datasets/SyMuPe/PianoCoRe}).
\end{itemize}

\section*{Acknowledgments}
\label{sec:acknowledgments}

The author would like to thank Vladimir Viro and Dmitrii Gavrilev for their feedback and suggestions regarding early versions of the alignment refinement algorithm and the dataset. The author is grateful to the TISMIR editorial team and the anonymous reviewers for their constructive and invaluable feedback, which improved the quality of the dataset and manuscript. 

The work was made possible by the use of the Zhores cluster and its computational resources \citep{zacharov2019zhores}. Furthermore, the author expresses gratitude to the creators of the MAESTRO, ASAP, (n)ASAP, ATEPP, GiantMIDI-Piano, Aria-MIDI, and PERiScoPe datasets. Their commitment to open science and the sharing of symbolic music resources provided the essential foundation for this work.

\section*{Competing Interests}
\label{sec:competing-interests}

The author has no competing interests to declare.

\section*{Author's Contribution}
\label{sec:contribution}

Ilya Borovik was responsible for the research conceptualization, methodology, software implementation, data curation, and the writing of the manuscript.

%%%%%%%%%%%%%%%%%%%%%%%%%%%%%%%%%%%%%%%%%%%%%%%%%%%%%%%%%%%%%%%%%%%%%%%%%%%%%%%%
% Please do not touch.
% Print Endnotes
\IfFileExists{\jobname.ent}{
   \theendnotes
}{
   %no endnotes
}

%%%%%%%%%%%%%%%%%%%%%%%%%%%%%%%%%%%%%%%%%%%%%%%%%%%%%%%%%%%%%%%%%%%%%%%%%%%%%%%%
% Bibliography
%%%%%%%%%%%%%%%%%%%%%%%%%%%%%%%%%%%%%%%%%%%%%%%%%%%%%%%%%%%%%%%%%%%%%%%%%%%%%%%%

% For bibtex users:
%%%TC:ignore
\bibliography{references}
%%%TC:endignore

\end{document}